\documentclass[a4paper, 12pt]{article} 
\usepackage{graphicx}
\usepackage{latexsym}

\newcommand{\be}{\begin{eqnarray}}
\newcommand{\ee}{\end{eqnarray}}
\begin{document}

\centerline{\bf Cosmology, Black Holes}
\centerline{and}
\centerline{\bf Shock Waves Beyond the Hubble Length}
\vspace{.3cm}

\centerline{December 18, 2002}

$$\begin{array}{ccc}
Joel\ Smoller\footnotemark[1]  &  Blake\ Temple\footnotemark[2] \nonumber \end{array}$$ 
\footnotetext[1]{Department of Mathematics, University of Michigan, Ann Arbor, MI 48109; Supported in
part by NSF Applied Mathematics Grant Number DMS-010-3998, and by the Institute of Theoretical Dynamics,
UC-Davis.}

\footnotetext[2]{Department of Mathematics, University of California, Davis, Davis CA
95616; Supported in part by NSF Applied Mathematics Grant Number DMS-010-2493, and by the Institute of
Theoretical Dynamics, UC-Davis.}

\newtheorem{Theorem}{Theorem}
\newtheorem{Lemma}{Lemma}
\newtheorem{Proposition}{Proposition}
\newtheorem{Corollary}{Corollary}
\small

\begin{abstract}

We construct exact, entropy satisfying shock wave solutions of the Einstein equations for a perfect fluid which extend the Oppeheimer-Snyder (OS) model
to the case of non-zero pressure, {\it inside the Black Hole}.  These solutions put forth a new Cosmological Model in which the expanding
Friedmann-Robertson-Walker (FRW) universe emerges from the Big Bang with a 
shock wave at the leading edge of the expansion, analogous to a classical shock wave explosion.   
This explosion  is large
enough to account for the enormous scale on which the galaxies and the
background radiation appear uniform.  In these models, the shock wave must lie beyond one Hubble length from the FRW
center, this threshhold being the
boundary across which the bounded mass lies inside its own Schwarzshild
radius, $2M/r>1,$ and thus the
shock wave solution evolves inside a Black Hole.  The entropy condition, which breaks the time symmetry, implies that the shock wave must weaken
until it eventually settles down to a zero pressure OS interface, bounding a {\em finite} total mass, that emerges from the White Hole event
horizon of an ambient Schwarzschild spacetime.   However, unlike shock matching outside a  Black
Hole, the equation
of state $p=\frac{c^2}{3}\rho,$ the equation of state at the earliest stage of Big
Bang physics, is {\em distinguished} at the
 instant of the Big Bang---for this equation of state alone, the shock wave
emerges from the Big Bang at a finite nonzero speed, the speed of light, decelerating to 
a subluminous wave from that time onward.  These shock wave solutions indicate a new cosmological model in which the Big Bang arises from a localized explosion
occurring inside the Black Hole of an asymptotically flat Schwarzschild spacetime.

\end{abstract}

\section{Introduction}
\setcounter{equation}{0}

In the standard model of cosmology based on a critically expanding, ($k=0$), Friedmann-Robertson-Walker (FRW) metric,
the universe is {\em infinite} at each instant after the Big Bang, \cite{peeb,smolte9,wein}. The
Hubble constant, which measures the recessional velocity of the galaxies, applies to the entire flat space ${\bf
R}^3$---that is, it applies to an entire universe of infinite mass and extent---at each fixed positive time in the standard model.  In this paper we present a
new cosmological model in which the expansion of the galaxies is a bounded expansion of finite total mass, and the Hubble law applies only to a bounded region
of spacetime, (c.f.
\cite{smolte9}). If the observed expansion of the galaxies actually only applies to a localized region of spacetime, then it follows that there must be a
wave at the leading edge of the expansion.  Thus to replace the assumption in the standard model that the universe is {\em infinite} at each instant
after the {\em Big Bang}, we are led to models in which the expansion  emerges from an event that is more similar to a classical shock wave explosion, than
it is to the usual scenario of the Big Bang.   

In previous work, we constructed such models by matching FRW metrics to standard Tolman-Oppenheimer-Volkoff
(TOV) metrics, (the metric for a static fluid sphere in general relativity), across a shock wave interface, \cite{smolte2,smolte9}.  In that work we derived
an upper limit on the distance that a shock wave could be from the FRW center, and this distance turned out to be closer than astronomical observations
suggest it could be.  (Astronomical observations currently extend out to approximately $1.5$ Hubble lengths.) In this paper we begin by showing that this
upper bound on the shock position that we identified in
\cite{smolte9} for standard FRW-TOV shock matching, is {\em exactly} equal to one Hubble length.  Indeed, we show that in order for the shock position to lie
beyond one Hubble length in an FRW metric, it is necessary that the spacetime beyond the shock wave lie {\em inside a Black Hole}.  Thus our previous shock
matching limit of one Hubble length results from the fact that a standard TOV metric cannot be continued into a Black Hole, except in the special case
when the pressure is zero, (we proved this in
\cite{smolte3}).  With this motivation, we here derive a new class of gravitational metrics that we call the TOV metric {\it inside the Black Hole}.  (In
contrast to the standard TOV metric, the TOV metric {\em inside the Black Hole} is dynamical.)  Based on this, we develop a theory of FRW-TOV shock
matching {\em inside the Black Hole}, and we use this to construct a new class of exact shock wave solutions of the Einstein equations, in which a shock wave
is incorporated into the FRW metric at distances arbitrarily far beyond the Hubble length.  This demonstrates that the limit in our previous work
\cite{smolte9}, that the shock position must lie within one Hubble length of the FRW center in FRW-TOV shock matching, can be overcome.  

In the exact solutions constructed in this paper, the
expanding FRW universe emerges behind a subluminous blast wave that explodes outward from the FRW origin
$\bar{r}=0$ at the instant of the Big Bang $t=0,$\footnotemark[3]
\footnotetext[3]{Here, $\bar{r}=R(t)r$ measures radial arclength distance at each fixed time $t$ in FRW coordinates, where $R(t)$ denotes the cosmological
scale factor, and $r$ is the standard FRW radial coordinate, c.f. (\ref{frw}), (\ref{rbar}) below.} 
 at a distance beyond one Hubble length\footnotemark[4]\footnotetext[4]{The Hubble length
$c/H(t)=\left[\frac{\dot{R}(t)}{cR(t)}\right]^{-1}$ depends on the cosmological scale factor $R(t),$ and an easy
calculation using the Einstein equations shows that in the FRW spacetime, the Hubble length increases with time.  Thus more and more galaxies
pass inside of the threshold distance of one Hubble length and come into view as time evolves.  
The Hubble length
$c/H_0$ at present time is estimated to be around $10^{10}$ light years.}.  The distance of one Hubble length is critical in the FRW metric because the
total FRW mass $M$ inside radius $\bar{r}$ satisfies $\frac{2M}{\bar{r}}<1,$ out to {\em exactly} one Hubble length.  Thus, one Hubble length marks the
event horizon of a Black Hole in a shock wave model in which the mass $M$ is isolated in an asymptotically Schwarzschild metric---the TOV metric in our
model.  (In contrast to the TOV metric outside the Black Hole, the TOV metric {\it inside the Black Hole} can be continued into an event
horizon, c.f. the remarks after Theorem \ref{theorem6} below.)   After the Big Bang, the shock wave in our exact solution continues to weaken as it expands
outward,  satisfying the entropy condition for shocks all the way out until the Hubble length eventually catches up to the shock wave
\footnotemark[5].\footnotetext[5]{In our cosmological interpretation of the FRW metric, we (loosely) identify the motion of the galaxies with the motion of
the FRW fluid, which is taken to be a perfect fluid with nonzero pressure, co-moving with the
FRW metric.    We show below that, although the shock wave moves {\em outward} through the galaxies,
($\dot{r}>0$), and the Hubble length increases with time, the number of Hubble lengths from the FRW center to the shock wave, (c.f. (\ref{Neqn})
below), as well as the total mass behind the shock wave, both {\em decrease} in time in these exact solutions, tending to infinity in backwards time at the
instant of the Big Bang.  This is no contradiction because the FRW pressure $p$ is assumed nonzero, c.f. Corollary 
\ref{corM} below.} At this instant the shock wave lies at the critical distance of exactly one Hubble length from the FRW center.
   From this time onward, the shock wave can be approximated by a zero
pressure, $k=0$ Oppenheimer-Snyder (OS) interface that emerges from the White Hole event horizon of an ambient Schwarzschild metric of finite
mass. (The entropy condition implies that the TOV density and pressure tend to zero as the shock interface approaches the critical distance of one Hubble
length.) Thereafter the interface continues out to infinity along a geodesic of the Schwarzschild metric outside the Black Hole.  Thus the OS solution gives
the large time asymptotics of this new class of shock wave solutions that evolve inside a Black Hole.

One of the surprises in the analysis is that the equation of
state that applies at the earliest stage of Big Bang physics,
$p=\frac{c^2}{3}\rho,$ is distinguished by the equations, and {\em only} for this equation of state does the shock wave emerge from the Big Bang at a finite
nonzero speed, the speed of light, decelerating to a subluminous wave from that time onward..  

These new shock wave solutions of the Einstein equations confirm the mathematical consistency of an FRW universe of finite extent and non-zero pressure
expanding outward from behind an entropy satisfying shock wave emerging from the origin at subluminous speed beyond one Hubble length at the instant of the
Big Bang, a prerequisite for early Big Bang physics.  Since the shock wave emerges from the Big Bang beyond one Hubble length, it would account for the
thermalization of radiation in a region that is initially well beyond the light cone of an observer positioned at the FRW center. Thus our attempt to incorporate a shock wave beyond one Hubble length has led to  unexpected and interesting connections between Big
Bang Cosmology and Black Holes.  But furthermore, we suggest that
general relativity pretty much forces the qualitative behavior we see here into any reasonable model that relaxes the
assumption in the standard model that the expansion of the galaxies is of
infinite extent at each fixed time.  (One could say that in these new models, the
{\it Copernican Principle} is replaced by the principle in physics that {\it Nothing Is
Infinite}.)

  
\vspace{.4cm}

In Section \ref{sect2} we transform the FRW metric to standard Schwarzschild coordinates, and use this to discuss the connection between the
Hubble length and the Schwarzschild radius.  In Section \ref{sect3} we construct the extension of the zero pressure, $k=0,$ OS solution to the interior of a
Black Hole by using Eddington-Finkelstein coordinates to regularize the event horizon of the Schwarzschild metric, \cite{misnthwh,smolte5}.  We 
return to these OS solutions in Section \ref{sect6} where we argue that the shock wave solutions constructed there continue naturally to the OS solution
after the solution has relaxed to almost zero pressure.  

In Section \ref{sect4} we construct the TOV metric {\it inside the Black Hole}.  We refer to this
metric as TOV because the metric components depend only on the radial coordinate
$\bar{r},$ but, as in the Schwarzschild metric, the TOV radial coordinate is timelike inside the Black Hole.  It follows that the mass function $M$ is
constant in each spacelike slice $\bar{r}=const.$ of the TOV metric {\it inside the Black Hole}. 

In Section \ref{sect5} we develop the theory of
shock matching between a
$k=0$ FRW metric and TOV metrics {\it inside the Black Hole}.  The shock matching techniques introduced in \cite{smolte1} must be modified inside the Black
Hole because the conservation constraint used in \cite{smolte1} introduces unphysical (characteristic) solutions.  The analysis leads to the derivation of a
system of differential equations that simultaneously describe the time evolution of the shock position together with the outer TOV pressure, and for
solutions of these equations, the shock interface must lie beyond one Hubble length from the center in the FRW metric.  One interesting feature of the
matching is that the radial coordinate
$\bar{r}$, a timelike coordinate in the TOV metric, is identified through shock matching with the FRW spacelike coordinate $\bar{r}$ that measures radial
arclength distance from the center at each fixed value of the (standard) time $t$ in the FRW metric.  Another interesting feature is that the mass function
$M,$ which is continuous across shocks as a consequence of shock matching, has the physical interpretation as a total mass inside radius $\bar{r}$ in the FRW
metric, but
$M$ has no such interpretation in the TOV metric by itself, and in fact we know of no general physical
interpretation of the mass function $M$ {\it inside the Black Hole}.

In Section \ref{sect6} we formulate the entropy condition, and construct a class of global, entropy satisfying shock wave solutions of the Einstein equations
under the simplifying assumption that the FRW sound speed $\sqrt{\sigma}$ is constant, (that is, we assume the FRW equation of state
$p=\sigma\rho,$
$\sigma=const.>0$, c.f.
\cite{smolte2}).  This includes the important case $\sigma=\frac{c^2}{3},$ usually assumed at the earliest stage of Big Bang physics.  (This is the equation
of state in the extreme relativistic limit of free particles, and for pure radiation, \cite{smolte1}.)  Remarkably, under a change of variables,
the shock matching equations of Section
\ref{sect5} reduce to a planar autonomous system when $\sigma=const.,$ and this system is amenable to global analysis, a requirement for the construction of
solutions in the large.  It is very interesting that the special value $\sigma=c^2/3$ plays distinguished role, and at this unique value, the
shock wave is everywhere subluminous after the Big Bang, but emerges from the Big Bang at exactly the speed of light.  For all other values of $\sigma,$ we
prove that the shock speed at the instant of the Big Bang is either zero, or infinite, and is everywhere subluminous if and only if $\sigma\leq c^2/3.$ This
is surprising because the equation of state
$p=\frac{c^2}{3}\rho$ played no special role in shock matching outside the Black Hole. 

The class of exact solutions in  Section
\ref{sect6} describes the global dynamics of an FRW universe of finite extent which explodes outward behind an entropy satisfying,
subluminous shock wave, emerging from the origin, beyond one Hubble length, at the instant of the Big Bang.  Because the TOV metric
{\it inside the  Black Hole}   has nonzero density and pressure, it follows that, unlike the OS solutions of Section \ref{sect3}, these new
exact shock wave solutions do not require any part of the empty space Schwarzschild solution inside the Black Hole for their construction.  
On the other hand, the fact that the OS solution gives the qualitative large time behavior of the solutions independent of $\sigma,$ (a consequence of the
entropy condition alone), suggests that the qualitative features of these solutions may be generic for a large class of equations of state.   

In Section
\ref{sect7} we obtain estimates for the shock position.  The conclusion of Theorem \ref{theorem9} is this: Let $t_0$ be the first time at which the shock
becomes visible at the FRW center.   In our exact solutions, we assume the FRW equation of
state 
$p=\sigma\rho,$
$\sigma=const,$ $0<\sigma\leq 1/3.$  (We use the convention that we take $c=1$ when convenient.)  For these solutions, the shock wave will first become
visible at the center
$\bar{r}=0$ of the FRW spacetime at FRW time
$t=t_0,$ at the moment when the Hubble length satisfies

\be
\frac{1}{H(t_0)}=\frac{1+3\sigma}{2}r_*,
\ee
where $r_*$ is the FRW position of the shock at the instant of the Big Bang.  (The actual arclength distance from the center to the shock at the Big
Bang will be $\bar{r}=0.$ )  At this time, the number of
Hubble lengths $\sqrt{N}_0$ from the FRW center to the shock wave at time $t=t_0$ satisfies the bounds  
\be
\nonumber
1\leq\frac{2}{1+3\sigma}\leq \sqrt{N}_0\leq \frac{2}{1+3\sigma}e^{\sqrt{3\sigma}\left(\frac{1+3\sigma}{1+\sigma}\right)}.
\ee
Thus, in particular, the shock wave will
still lie beyond one Hubble length
at the FRW time $t_0$ when it first becomes visible.  Furthermore,the time $t_{crit}>t_0$ at which the shock wave will emerge from the
White Hole event horizon, given that $t_0$ is the first instant at which the shock becomes visible at the FRW center, can be estimated by

\be
\nonumber
\frac{2}{1+3\sigma}e^{\frac{1}{4}\sigma}\leq \frac{t_{crit}}{t_0}\leq
\frac{2}{1+3\sigma}e^{\frac{2\sqrt{3\sigma}}{1+\sigma}},
\ee
and by the better estimate
\be
\nonumber
e^{\frac{\sqrt{6}}{4}}\leq \frac{t_{crit}}{t_0}\leq
e^{\frac{3}{2}},
\ee
in the case $\sigma=1/3.$
For example, (\ref{finalestimate1}), (\ref{finalestimate2-22}) imply that at the OS limit $\sigma=0,$ 
$$\sqrt{N_0}=2,\ \  \frac{t_{crit}}{t_0}=2,$$ 
and in the limit $\sigma=1/3,$
$$1\leq\sqrt{N_0}\leq 1.95,\ \ 1.8\leq\frac{t_{crit}}{t_0}\leq 4.5.$$

We conclude in these shock wave cosmological models, that at the moment $t_*$ when the shock wave first becomes visible at the
FRW center, it must lie within 4.5 Hubble lengths of the center.  Throughout the expansion up until this time, the expanding
universe must lie entirely within a {\em Black Hole}---the universe will eventually emerge from this {\em Black Hole}, but not until some later time
$t_{crit},$ where $t_{crit}$ does not exceed $3t_*.$  

Concluding remarks are made in the final section.

\section{The Hubble Distance and the Schwarzschild Radius}
\setcounter{equation}{0}\label{sect2}

According to Einstein's Theory of General Relativity, all properties of the gravitational field are determined by
a Lorentzian spacetime metric tensor $g,$ whose line element in a given coordinate system $x=(x^0,...,x^3)$  is
given by

\begin{equation}
ds^2=g_{ij}dx^idx^j.
\label{q1.2}
\end{equation}
(We use the Einstein summation convention whereby {\em all repeated up-down indices are assumed summed from $0$
to
$3.$}) The components $g_{ij}$ of the gravitational metric $g$ satisfy the Einstein equations,

\begin{equation}
G^{ij}=\kappa T^{ij},\ \ \ T^{ij}=(\rho c^2+p)w^iw^j+pg^{ij},
\label{einstein}
\end{equation}
where we assume the stress-energy tensor $T$ of a perfect fluid.  Here $G$ is the
Einstein curvature tensor,

\begin{equation}
\kappa=\frac{8\pi{\cal G}}{c^4}
\label{kappa}
\end{equation} 
is the coupling constant, ${\cal G}$ is Newton's gravitational constant, $c$
is the speed of light,
$\rho c^2$ is the energy density, $p$ is the pressure, and ${\bf w}=(w^0,...,w^3)$ are the components of the
$4$-velocity of the fluid, c.f. \cite{wein}.  We use the convention that we take $c=1$ and ${\cal G}=1$ when
convenient.

In this section we consider the FRW metric, a spacetime metric whose line element takes the form

\begin{equation}
ds^2=-dt^2+R(t)^2\left(\frac{dr^2}{1-kr^2}+r^2d\Omega^2\right),
\label{frw}
\end{equation}
where $R(t)$ is the so-called cosmological scale factor, and $d\Omega^2=d\theta^2+sin^2\theta d\phi^2$ denotes the line element on the unit 2-sphere.
The FRW metric describes the time evolution of a three dimensional space of constant scalar curvature, 
(the $t$=const. surfaces),
and $k$ can be assumed to take one of the 
values $\{-1,0,1\}$ via a rescaling of the
radial coordinate $r.$ Radial distance at each
fixed time in the FRW metric is measured by $\bar{r},$ where 

\begin{equation}
\bar{r}=R(t)r,
\label{rbar}
\end{equation}
and it is standard to rescale time so that $R$ goes from $0$ to $1,$ where $R=0,$ $t=0$ corresponds to the Big Bang,
and $R=1,$ $t=t_0$ corresponds to present time.

 The Hubble constant $H,$ (which actually depends on $t$), is given in terms of $R(t)$ by

\begin{equation}
\label{hubble}
H\equiv H(t)=\frac{\dot{R}(t)}{R(t)},
\end{equation}
and we let $H_0=H(t_0)$ 
denote the present value of the Hubble constant.  

The Hubble length $c/H$, (the reciprocal of the Hubble constant when we take $c=1$), gives a length scale determined by the
expansion rate of the galaxies, and can be interpreted as the travel distance for
a light ray starting at the Big Bang and ending at time $t.$ (That is, the age of the universe is approximately
$H_0^{-1}\approx 10^{10}$ years, so the Hubble length $c/H_0\approx 10^{10}$ light years, is a measure of the distance light travels during
this time interval, \cite{wein}.)   The Hubble length is thus a measure of the distance to the furthest objects that can be seen in the universe at a given
time.  (Estimates for this distance in terms of radial arclength $\bar{r}$ at fixed FRW time $t$ are given in Theorems 
\ref{thmrinfty} and \ref{thmrinfty-1} below.)

Putting the metric ansatz (\ref{frw}) into the Einstein equations
(\ref{einstein}) gives the equations for the FRW metric, \cite{wein},

\begin{equation}
\label{frw1}
H^2=\left(\frac{\dot{R}}{R}\right)^2=\frac{\kappa}{3}\rho-\frac{k}{R^2},
\end{equation}
and

\begin{equation}
\label{frw2}
\dot{\rho}=-3(p+\rho)H.
\end{equation}
The unknowns $R,$ $\rho$ and $p$ are 
assumed to be functions of the FRW coordinate time $t$ alone, and \lq\lq dot''  
denotes 
differentiation with respect to $t.$   In this paper, we focus on the case of critical expansion,
the case
$k=0,$
\cite{blaugu,wein}.  

When $k=0,$ the $t=const.$ surfaces in the FRW metric are infinite, flat, Euclidean $3$-space.  It follows
that the standard model of cosmology based on a critically expanding FRW metric, \cite{blaugu}, implicitly assumes that the expansion rate $H$ of the
universe is constant all the way out to infinity at each fixed time $t$ in the FRW metric---this even though, based on lookback time, we can in principle
only observe the universe out to about $10^{10}$ light years. 

If, on the other hand, there is a shock wave at the leading edge of the
expansion of the galaxies, then the following question presents itself:  What is the critical radius
$\bar{r}_{crit}$ at each fixed time $t$ in a $k=0$ FRW metric such that the total mass inside a shock wave positioned beyond that radius puts the universe
inside a Black Hole?   Indeed, there must always be such a critical radius because the total mass
$M(\bar{r},t)$ inside radius
$\bar{r}$ in the FRW metric at fixed time $t$ increases like $\bar{r}^3,$ and so at each fixed time $t$ we must have 
$\bar{r}>2M(\bar{r},t)$ for small enough
$\bar{r},$ while the reverse inequality holds for large $\bar{r}.$ Thus for every time $t$ there must exist a smallest
$\bar{r}=\bar{r}_{crit}$ for which
$\bar{r}_{crit}=2M(\bar{r}_{crit},t).$ This critical radius then marks the Schwarzschild
radius in the metric that lies beyond the shock wave when it is at that position.  We will presently show that when $k=0,$ $\bar{r}_{crit}$ equals
the Hubble length, and this explains the shock matching limit uncovered in our previous work in terms
of the Hubble length, (c.f. Section
6, \cite{smolte9}). That is, we cannot match a critically
expanding FRW metric to a classical TOV metric beyond one Hubble length without continuing the TOV solution into a Black Hole---and we showed in
\cite{smolte3} that the standard TOV metric cannot be continued into a Black Hole.  

At this point we make a note on terminology: we say that a radial solution of the Einstein equations is {\it inside
the Black Hole} provided that $\frac{2M}{\bar{r}}>1.$ If further, the radial coordinate $\bar{r}$ is always
increasing along timelike curves, then we also refer to such a region $\frac{2M}{\bar{r}}>1$ as a {\it White Hole}, but we will nevertheless use the term {\it
Black Hole} whenever $\frac{2M}{\bar{r}}>1.$. 
For example, in the Kruskal development of a Schwarzschild Black Hole there are four quadrants, and one quadrant contains the {\it Black
Hole singularity} and the opposite quadrant, (its time reversal), contains the {\it White Hole singularity},
\cite{misnthwh}.  Both quadrants lie within the region where $\frac{2M}{\bar{r}}>1,$ and the quadrant with the Black [resp. White] Hole singularity at
$\bar{r}=0$ has the property that all timelike geodesics end [resp. begin] at the singularity in finite
proper time\footnotemark[6].
\footnotetext[6]{By itself, Einstein's theory is {\it time reversible}, the direction of forward time is left undetermined, and extra conditions like
entropy conditions for shocks are needed to determine the direction of forward time. In the region $2M/\bar{r}>1$ outside the event horizon, the
Schwarzschild metric is symmetric under time reversal in the sense that the time reversal of a geodesic remains a geodesic. However in the region
$2M/\bar{r}<1,$ the time reversal of the Schwarzschild metric takes the Black Hole region beyond $t=+\infty$ to the White Hole region before $t=-\infty$.
Thus without extra conditions, the expanding White Hole is just as admissible a solution of the Einstein equations as is its time reversal, the collapsing
Black Hole.} Thus if
$\bar{r}_{crit}=c/H,$ and the mass function
$M$ is continuous across shock waves, (both of which we demonstrate below), then it follows that one can incorporate a shock wave into an FRW metric beyond
one Hubble length only if the metric beyond the shock wave lies
{\it  inside a Black Hole}, where $\frac{2M(\bar{r},t)}{\bar{r}}>1.$
\vspace{.2cm} 

We now verify that $\bar{r}_{crit}=c/H,$ and that
the Hubble length is the limit for FRW-TOV shock matching outside a Black Hole, c.f. \cite{smolte9}. 
To this end, write the FRW metric (\ref{frw}) in standard Schwarzschild coordinates ${\bf \bar{x}}=(\bar{r},\bar{t})$ where the
metric takes the form

\begin{equation}
\label{sschwarz}
ds^2=-B(\bar{r},\bar{t})d\bar{t}^2+A(\bar{r},\bar{t})^{-1}d\bar{r}^2+\bar{r}^2d\Omega^2,
\end{equation}
and the mass function $M(\bar{r},\bar{t})$ is defined through the relation
\be
A=1-\frac{2M}{\bar{r}}.
\label{M}
\ee
(We have set ${\cal G}=c=1,$ and our notation is to denote the standard, non-angular FRW coordinates by ${\bf x}=(x^0,x^1)=(t,r),$ but
TOV coordinates are denoted by
${\bf \bar{x}}=(\bar{x}^0,\bar{x}^1)=(\bar{r},\bar{t})$ because we work inside the Black Hole where $A<0$ and $\bar{r}$ is the timelike
coordinate.) It is well known that a general spherically symmetric metric can be transformed to the form (\ref{sschwarz}) by coordinate
transformation, \cite{wein}.  To obtain $A$ and $B$ for the FRW metric (\ref{frw}), set
$\bar{r}=Rr,$ so that

$$
dr=\frac{d\bar{r}}{R}-\frac{\bar{r}}{R^2}\dot{R}dt.
$$
Using this in the FRW line element we find

\begin{equation}
\label{frwdrdt}
ds^2=-Cdt^2+Dd\bar{r}^2+2Ed\bar{r}dt+\bar{r}^2d\Omega^2,
\end{equation}
where

\be
C&=&\frac{1-kr^2-H^2\bar{r}^2}{1-kr^2},\label{C}\\
D&=&\frac{1}{1-kr^2},\label{D}\\
E&=&-\frac{H\bar{r}}{1-kr^2}.\label{E}
\ee
Now define the time coordinate $\bar{t}=\bar{t}(t,\bar{r})$ by

\be
d\bar{t}=(\psi C)dt-(\psi E)d\bar{r},
\label{exactfirst}
\ee
where $\psi$ is an integrating factor to be determined by the equation
\begin{equation}
\label{frwschwarz}
\frac{\partial}{\partial \bar{r}}(\psi C)+\frac{\partial}{\partial t}(\psi E)=0.
\end{equation}
Equation (\ref{frwschwarz}) implies that $d\bar{t}$ is an exact differential, and the choice of $E$ and $C$ applied to
(\ref{exactfirst}) removes the cross term from the metric (\ref{frwdrdt}).  In $(\bar{r},\bar{t})$ coordinates the FRW
metric takes the form

\begin{equation}
\label{frwschwarz00}
ds^2=-\frac{1}{C\psi^2}d\bar{t}^2+
\left(D+\frac{E^2}{C}\right)d\bar{r}^2+\bar{r}^2d\Omega^2.
\end{equation}
Using (\ref{C})-(\ref{E}) in (\ref{frwschwarz00}) gives

\begin{equation}
\label{frwschwarz1}
ds^2=-\frac{1}{\psi^2}\left\{\frac{1-kr^2}{1-kr^2-H^2\bar{r}^2}\right\}d\bar{t}^2+
\left\{\frac{1}{1-kr^2-H^2\bar{r}^2}\right\}d\bar{r}^2+\bar{r}^2d\Omega^2.
\end{equation}

Comparing (\ref{frwschwarz1}) with (\ref{sschwarz}) we obtain

\be
A^{-1}&=&\frac{1}{1-kr^2-\left(\bar{r}H\right)^2},\label{A-1}\\
B&=&\frac{1}{\psi^2}\left(\frac{1-kr^2}{1-kr^2-\left(\bar{r}H\right)^2}\right).\label{B-1}
\ee
Note that the characteristic curves for (\ref{frwschwarz}) are given by

\be
\frac{d\bar{r}}{dt}=\frac{C}{E}=\frac{-1+kr^2+H^2\bar{r}^2}{H\bar{r}},
\label{char}
\ee
the RHS being a smooth function for all $t>0$ for which $H\neq0\neq\bar{r}.$  (By (\ref{frw1}), the condition
$H=\frac{\dot{R}}{R}\neq0$ holds for all
$t>0$ when $k=0,-1,$ and it holds except at the turning point $\dot{R}=0$ in the case
when
$k=+1,$ c.f.
\cite{wein}.)  It follows that a solution
$\psi$ of (\ref{frwschwarz}) exists in a neighborhood of any point.  In fact, we have more:  we can assign arbitrary
initial values for $\psi$ on any surface that is non-characteristic for (\ref{frwschwarz}),  and then use these
values to solve (\ref{frwschwarz}) for $\psi$ in a neighborhood of that surface, c.f.
\cite{smolte2,wein}.  By (\ref{B-1}), this implies that we can assign arbitrary (non-negative) values for $B$ on any
non-characteristic surface.    We conclude that (\ref{frwschwarz1})-(\ref{B-1}) gives the FRW metric in standard Schwarzschild form, the
coefficient of
$d\bar{r}^2$ is determined independently of $\psi,$ and values of $B$ can be freely assigned (locally) on any smooth non-characteristic surface in
$(t,\bar{r})$ space, thus determining initial values for equation (\ref{frwschwarz}).

Moreover, using (\ref{A-1}) in (\ref{M}), it follows that

\be
M(t,\bar{r})&=&\frac{\bar{r}}{2}(1-A)=\frac{\bar{r}}{2}\left(\bar{r}^2H^2\right)=
\frac{1}{3}\frac{\kappa}{2}\rho\bar{r}^3\nonumber\\
&=&\frac{\kappa}{2}\int_0^{\bar{r}}\rho(t)s^2ds.
\label{M2}
\ee
Since in the FRW metric $\bar{r}=Rr$ measures arclength along radial geodesics at fixed time, we see from (\ref{M2}) that
$M(t,\bar{r})$ has the physical interpretation as the total mass inside radius $\bar{r}$ at time $t$ in the FRW metric. 

From here on, we restrict to the case of critical expansion,
the case
$k=0.$
In this case, since $\frac{2M}{\bar{r}}=1-A,$ one sees from (\ref{A-1}) that $\bar{r}=H^{-1}$ is equivalent to
$\frac{2M}{\bar{r}}=1,$ and so the following
equivalences are valid at any fixed time
$t$:

\be
\bar{r}=H^{-1}\ \ {\rm iff}\ \ \frac{2M}{\bar{r}}=1\ \ {\rm iff}\ \ A=0.
\label{hubble3}
\ee
We conclude that
$
\bar{r}=H^{-1}
$
is the critical length scale for the FRW metric at fixed time $t$ in the sense that
$\frac{2M}{\bar{r}}-1$ changes sign at $\bar{r}=H^{-1},$ and so the universe lies {\it inside a Black Hole} beyond
$\bar{r}=H^{-1},$ as claimed above.  

From the FRW equation (\ref{frw1}), it also follows that 
$\bar{r}=H^{-1}$ iff 
\begin{equation}
\label{hubble4}
r^2=\frac{3}{\kappa}\frac{1}{\rho R^2}.
\end{equation}
This explains the condition

\begin{equation}
\label{hubble5}
r^2\leq\frac{3}{\kappa}\frac{1}{\rho R^2},
\end{equation}
 given in equation (6.2) of \cite{smolte9} for the maximal position of a shock wave in FRW-TOV matching.  This condition
was derived in \cite{smolte9} from a different point of view.  Thus (\ref{hubble3}), (\ref{hubble4}) provides a physical
interpretation for the bound (6.2) in
\cite{smolte9}; namely, that the shock wave lies inside the
Hubble length. 

The Hubble length $\bar{r}_{crit}=\frac{c}{H}$ is also the critical distance at which the outward expansion of the FRW metric exactly cancels the inward
advance of a radial light ray impinging on an observer positioned at the origin of a $k=0$ FRW metric. Indeed, by (\ref{frw}), a light ray traveling radially
inward toward the center of an FRW coordinate system satisfies,
\be
\label{lightray1}
c^2dt^2=R^2dr^2,
\ee
so that

\begin{equation}
\label{lightray}
\frac{d\bar{r}}{dt}=\dot{R}r+R\dot{r}=H\bar{r}-c=H(\bar{r}-\frac{c}{H})>0,
\end{equation}
if and only if

$$
\bar{r}>\frac{c}{H}.
$$
Thus the arclength distance from the origin to an inward moving light ray at fixed time $t$ in a $k=0$ FRW metric will actually {\em increase} as long
as the light ray lies beyond the Hubble length.  An inward moving light ray will, however, eventually
cross the Hubble length and reach the origin in finite proper time, due to the increase in the Hubble length with time.  

We now calculate the infinite redshift limit in terms of the Hubble length.  It is well known that light emitted at $(t_e,r_e)$ at wavelength $\lambda_e$
in an FRW spacetime will be observed at $(t_0,r_0)$ at wavelength $\lambda_0$ if

$$
\frac{R_0}{R_e}=\frac{\lambda_0}{\lambda_e}.
$$
Moreover, the redshift factor $z$ is defined by
$$
z=\frac{\lambda_0}{\lambda_e}-1.
$$
Thus, infinite redshifting occurs in the limit $R_e\rightarrow0,$ where $R=0,$ $t=0$ is the Big Bang.  

Consider now a light ray emitted at the instant of the Big Bang, and observed at the FRW origin at present time $t=t_0.$  Let $r_{\infty}$ denote the FRW
coordinate at time $t\rightarrow 0$ of the furthestmost objects that can be observed at the FRW origin before time
$t=t_0.$   Then $r_{\infty}$ marks the position of objects at time $t=0$ whose radiation would be observed as infinitly redshifted, (assuming no
scattering).   Note then that a shock wave emanating from $\bar{r}=0$ at the instant of the Big Bang, will be observed at the FRW origin before present time
$t=t_0$ only if its position $r$ at the instant of the Big Bang satisfies $r< r_{\infty}.$  We now estimate $r_{\infty}.$ 

First, from (\ref{lightray1}) it follows that an incoming radial light ray in an FRW metric follows a lightlike trajectory $r=r(t)$ if

$$
r-r_e=-\int_{t_e}^{t}\frac{d\tau}{R(\tau)},
$$
and thus

\begin{equation}
r_{\infty}=\int_{0}^{t_0}\frac{d\tau}{R(\tau)}.
\label{rinfty}
\end{equation}
We now prove the following theorem:

\begin{Theorem}
\label{thmrinfty}
If the pressure $p$ satisfies the bounds

\begin{equation}
0\leq p\leq\frac{1}{3}\rho,
\label{bounds2}
\end{equation}
then for any equation of state, the age of the universe $t_0$ and the infinite red shift limit $r_{\infty}$ are bounded in terms of the Hubble length by
\begin{equation}
\frac{1}{2H_0}\leq t_0\leq\frac{2}{3H_0},
\label{bounds4}
\end{equation}

\begin{equation}
\frac{1}{H_0}\leq r_{\infty}\leq \frac{2}{H_0}.
\label{rinfinitybound}
\end{equation}
\end{Theorem}
(We have assumed that $R=0$ when $t=0$ and $R=1$ when $t=t_0$ and $H=H_0.$  Note that $p=\frac{1}{3}\rho$ is the extreme relativistic limit of free
particles, as well as the equation of state for pure radiation,
\cite{wein}.)
\vspace{.2cm}

\noindent{\bf Proof:}  Integrating (\ref{frw1}) we obtain

\begin{equation}
R(t)=e^{-\int_{t}^{t_0}\sqrt{\frac{\kappa\rho}{3}}dt},
\label{intfrw1}
\end{equation}
and, from (\ref{frw2}) we have

\begin{equation}
-\int_{\infty}^{\rho}\frac{d\rho}{(p+\rho)\sqrt{\rho}}=\sqrt{3\kappa}t.
\label{intfrw2}
\end{equation}
Thus, if we know $p$ as a function of $\rho,$ then we can integrate (\ref{intfrw2}) to obtain $\rho$ as a function of $t,$ and then use this in
(\ref{intfrw1}), which can then be used in (\ref{rinfty}) to calculate $r_{\infty}.$  Assuming (\ref{bounds2}), we can estimate (\ref{intfrw2}) by

\begin{equation}
-\int_{\infty}^{\rho}\frac{d\rho}{\frac{4}{3}\rho^{3/2}}\leq\sqrt{3\kappa}t\leq-\int_{\infty}^{\rho}\frac{d\rho}{\rho^{3/2}},
\nonumber
\end{equation}
which leads to

\begin{equation}
\frac{3}{4\kappa t^2}\leq\rho\leq\frac{4}{3\kappa t^2}.
\label{bounds2-1}
\end{equation}
Since (\ref{frw1}) is $H^2=\frac{\kappa}{3}\rho$ when $k=0,$ (\ref{bounds2-1}) gives

\begin{equation}
\frac{1}{2t}\leq H\leq\frac{2}{3t},
\label{bounds3}
\end{equation}
which implies (\ref{bounds4}). To estimate $R(t),$ write (\ref{bounds3}) in the form

\begin{equation}
\frac{1}{2t}\leq \frac{d}{dt}ln R\leq\frac{2}{3t},
\nonumber
\end{equation}
which integrates to

\begin{equation}
\left(\frac{t}{t_0}\right)^{\frac{2}{3}}\leq R\leq\left(\frac{t}{t_0}\right)^{\frac{1}{2}}.
\label{bounds5}
\end{equation}
Finally, using (\ref{bounds5}) in (\ref{rinfty}) gives

\begin{equation}
\int_{0}^{t_0}\left(\frac{t_0}{t}\right)^{1/2}dt\leq r_{\infty}\leq\int_{0}^{t_0}\left(\frac{t_0}{t}\right)^{2/3}dt,
\nonumber
\end{equation}
which leads directly to (\ref{rinfinitybound}).  $\Box$

The next theorem gives closed form solutions of the FRW equations (\ref{frw1}), (\ref{frw2}) in the case when the sound speed
$\sqrt{\sigma}\equiv constant.$ 
These solutions are the starting point of the exact shock wave solutions constructed in Section 6.   As a special case we recover
the bounds in (\ref{bounds4}) and (\ref{rinfinitybound}) from the cases
$\sigma=0$ and $1/3.$

\begin{Theorem}
\label{thmrinfty-1}\label{theorem2}
Assume $k=0$ and the equation of state

\begin{equation}
p=\sigma\rho,
\nonumber
\end{equation}
where $\sigma$ is assumed constant,
$$
0\leq\sigma\leq1.
$$
Then, (assuming
an expanding universe $\dot{R}>0$), the solution of system (\ref{frw1}), (\ref{frw2}) satisfying $R=0$ at $t=0$ and $R=1$ at $t=t_0$ is given by,
\be
&\rho=\frac{4}{3\kappa(1+\sigma)^2}\frac{1}{t^2},&\label{frw3-1}\\
 &R=\left(\frac{t}{t_0}\right)^{\frac{2}{3(1+\sigma)}},&\label{frw4-1}\\
&\frac{H}{H_0}=\frac{t_0}{t}.&\label{frwH}
\ee
Moreover, the age of the universe $t_0$ and the infinite red shift limit $r_{\infty}$ are given exactly in terms of the Hubble length by

\be
t_0=\frac{2}{3(1+\sigma)}\frac{1}{H_0},\label{sigmaage}
\ee

\begin{equation}
r_{\infty}=\frac{2}{1+3\sigma}\frac{1}{H_0}.\ \ \ \ 
\label{inftyexact}
\end{equation} 
\end{Theorem}
From (\ref{inftyexact}) we conclude that a shock wave will be observed at the FRW origin before
present time $t=t_0$ only if its position $r$ at the instant of the Big Bang satisfies $r<\frac{2}{1+3\sigma}\frac{1}{H_0}.$  Note that $r_{\infty}$
ranges from one half to two Hubble lengths as $\sigma$ ranges from $1$ to $0,$  taking the intermediate value of one Hubble length at
$\sigma=1/3,$ c.f. (\ref{rinfinitybound}).
\vspace{.2cm}

\noindent{\bf Proof:}   Formulas (\ref{frw3-1})-(\ref{frwH}) follow directly from (\ref{intfrw1})-(\ref{intfrw2}), and agree with the
formulas given in \cite{smolte2}.  Differentiating (\ref{frw4-1}) at $t=t_0$ gives (\ref{sigmaage}), and using (\ref{frw4-1}) and (\ref{sigmaage}) in
(\ref{rinfty}) gives (\ref{inftyexact}).  $\Box$

\begin{Corollary}\label{corM}
If $p=\sigma\rho,$ $\sigma=const.>0,$  then the total mass inside radius $r=const,$ (that is, inside a ball whose boundary is comoving with
the galaxies), decreases in time.
\end{Corollary}

\noindent{\bf Proof:}  Using (\ref{frw3-1})-(\ref{frw4-1}) in (\ref{M2}), it follows that

\be
M=\frac{\kappa}{2}\int_0^{\bar{r}}\rho(t)s^2ds
=\frac{2\bar{r}^3}{9(1+\sigma)^2t_0^{\frac{2}{1+\sigma}}}t^{\frac{-2\sigma}{1+\sigma}},\label{Mdotneg}
\ee
so $\dot{M}<0$ if $\sigma>0,$

\section{The OS Solution Inside the Black Hole}
\setcounter{equation}{0}\label{sect3}

The simplest model of a localized FRW metric contained within a shock boundary that lies beyond the Hubble distance, is
one in which
$p\equiv0,$ and the FRW metric is matched to the Schwarzschild metric at a contact discontinuity interface positioned
{\it inside the Black Hole} of the Schwarzschild metric.  This poses the problem of extending the OS solution
smoothly into the interior of a Schwarzschild Black Hole.  In this section we construct the OS solution inside the Black
Hole in the case $k=0,$ making the point of discussing it from the FRW point of view, in terms of the Hubble length.   (See \cite{oppesn} and \cite{misnthwh}
for a discussion of the case $k<0.$)  The $k=0$ OS solution inside the Black Hole will give us a point of comparison for the
$p\neq0$ shock wave models constructed in Sections 5 and 6, and we argue in Section 7 that the (expanding) OS solution also describes the large time
asymptotics of these shock wave models.   When
$p\neq0,$ the resulting interface is a shock wave, and an entropy condition breaks the time symmetry.  In our examples of Section 6, it is the {\em outward}
expanding solution (explosion) that globally meets the entropy condition, (not the inward collapse), thus making the cosmological interpretation of
the model relevant.  In contrast, the interface in the OS solution is {\it time reversible}, but the collapsing solution is the one that is relevant in the
standard interpretation of the OS solution as a pressureless sphere collapsing into a Black Hole, \cite{oppesn,misnthwh}. 

Our method is to match the $k=0$ FRW metric to the empty space Schwarzschild metric written in Eddington-Finkelstein (EF) coordinates
\cite{misnthwh}, across a shock interface\footnotemark[7]\footnotetext[7]{The interface in the OS solutions is, in the language of gas dynamics, a contact
discontinuity, which is time reversible because neither mass nor momentum cross the interface.} that lies beyond the Hubble length on the FRW side of the
shock. As in a classical explosion, we assume that the FRW metric lies inside a bounded region behind an outgoing shock interface, and for the OS solution we
assume that the Schwarzschild metric describes the spacetime beyond the interface.  Thus in the OS model,  the shock wave (contact discontinuity) marks the
leading edge of the expansion of the FRW metric.  

The EF
coordinates provide a regularization of the Schwarzschild metric at the event horizon, and the mapping from the EF
spacetime to the Schwarzschild spacetime is a 1-1 mapping, \cite{misnthwh}.  Both coordinate systems cover the region of the
Schwarzschild metric outside the event horizon, together with that portion of the Black Hole that is coordinatized by
the Schwarzschild coordinates.  Here, since we are dealing with an expanding FRW solution, we work with the time
reversal of these metrics, which covers the region beyond the event horizon, together
with the White Hole region
{\it inside the Black Hole}  We show
that with this latter time orientation, we can match the EF metric to the spacetime described by a critically {\em expanding}
FRW metric across a discontinuity at finite radius, and its time reversal provides the corresponding matching to
the critically {\em contracting} FRW metric.

We return to these OS solutions in Section 6 where we argue that
the shock wave solutions constructed there continue naturally to the OS solution after the TOV density and pressure tend to zero, assuming that
the shock has relaxed to a sufficiently weak wave.    But we find this OS model interesting in its own right because it not only provides the
simplest model in which the expansion of the galaxies corresponds to the expansion of a finite total mass with a wave at the leading edge of the
expansion, but it also embeds the Big Bang singularity of an FRW metric within the singularity of a larger spacetime---the larger spacetime being
the empty space Schwarzschild solution, and its singularity being the White Hole singularity that lies inside the event horizon in the Kruskal
development of the Schwarzschild metric.  Such an embedding is possible {\em only} under the assumption that the Hubble law applies to a {\em
bounded} region of spacetime, because the infinite FRW metric cannot be matched to the Schwarzschild metric.

In EF coordinates, the Schwarzschild line element takes the form

\be
d\hat{s}^2=-Ad\hat{t}^2-2d\bar{r}d\hat{t}+\bar{r}^2d\Omega^2,
\label{EF}
\ee
where

\be
A=1-\frac{2M}{\bar{r}},
\label{A}
\ee
and $M$ is the constant mass of the Schwarzschild solution.  Indeed, using (\ref{frwdrdt})-(\ref{frwschwarz}), the time
coordinate $\bar{t}$ that eliminates the mixed term in (\ref{EF}) satisfies

\be
\label{EF-1}
d\bar{t}=\psi Ad\hat{t}-\psi\bar{r},
\ee
where $\psi$ satisfies

\be
\label{EF-2}
\frac{\partial}{\partial\hat{t}}\psi+\frac{\partial}{\partial\bar{r}}(\psi A)=0.
\ee
The solution to (\ref{EF-1}), (\ref{EF-2}) is

\be
\label{EF-3}
\psi=1/A,
\ee

\be
\label{EF-4}
\bar{t}=t-\bar{r}-ln|\bar{r}-2GM|,
\ee
and in $(\bar{r},\bar{t})$-coordinates the metric takes the standard Schwarzschild form 

\be
\label{EF-5}
ds^2=-Ad\bar{t}^2+\frac{1}{A}d\bar{r}^2+\bar{r}^2d\Omega^2.
\ee
Thus (\ref{EF-4}) defines a $1-1$ mapping
$(\hat{t},\bar{r})\rightarrow(\bar{r},\bar{t})$ taking $\left\{\bar{r}>2M\right\}\times{\bf R}$ to itself, and
another
$1-1$ mapping $(\hat{t},\bar{r})\rightarrow(\bar{r},\bar{t})$ taking
$\left\{0<\bar{r}<2M\right\}\times{\bf R}$ to itself.  This verifies the claim that the
EF metric covers two of the four quadrants of the (maximal) Kruskal development of the Schwarzschild metric in a $1-1$
fashion.  Unlike the Schwarzschild metric, the EF metric is smooth at $\bar{r}=2M,$ and this is reflected in
the singularity of the transformation (\ref{EF-4}) at $\bar{r}=2M.$  Note that in
$(\bar{r},\bar{t})$ coordinates, the vector field $\frac{\partial}{\partial\bar{r}}$ is timelike, spacelike,
when $\bar{r}<2M,$ $\bar{r}>2M,$ respectively; while in
$(\hat{t},\bar{r})$ coordinates, the vector field $\frac{\partial}{\partial\bar{r}}$ is lightlike.  This is no
contradiction because $\frac{\partial}{\partial\bar{r}}$ points along the level curves of its complementary coordinates in
a given coordinate system.  On the other hand, the vector fields
$\frac{\partial}{\partial\bar{t}}$ and
$\frac{\partial}{\partial\hat{t}}$ both point along the level curves of
$\bar{r},$ and hence retain the same character as spacelike or timelike according to the sign of $A.$

The
$k=0$ FRW metric is given in (\ref{frw}) as

\be
ds^2=-dt^2+R(t)^2dr^2+R(t)^2r^2d\Omega^2,
\label{FRW0}
\ee
where $R(t)$ is the cosmological scale factor.  To obtain a matching of the two metrics (\ref{EF}), (\ref{FRW0}), that is
smooth at the Schwarzschild radius, we match them in $(\hat{t},\bar{r})$ coordinates.  To start, first match the spheres of
symmetry by setting, (c.f.
\cite{smolte1,wein}),
$$
\bar{r}=Rr.
$$
Writing (\ref{frw}) in $(t,\bar{r})$ coordinates yields the following form of the FRW metric derived
in (\ref{frwdrdt})-(\ref{E}) above upon setting $k=0$:

\begin{eqnarray}
ds^2&=&-dt^2+\left\{d\bar{r}^2-2H\bar{r}d\bar{r}dt+\frac{\dot{R}^2}{R^2}\bar{r}^2dt^2\right\}+\bar{r}^2d\Omega^2.
\nonumber\\
\label{frw-1}
\end{eqnarray}
We now construct the transformation $\hat{t}=\hat{t}(t,\bar{r})$ by finding functions $b(t,\bar{r})$ and
$\phi(t,\bar{r})$ such that $b$ and $\phi$ satisfy

\be
d\hat{t}=\phi dt+\phi bd\bar{r},
\label{dhat}
\ee
and

\be
\phi_{\bar{r}}-\left(\phi b\right)_t=0.
\label{exact}
\ee
Equation (\ref{exact}) implies that $d\hat{t}$ is an exact differential, and thus (\ref{dhat}) defines the coordinate
transformation for $\hat{t}$ as a function of $t$ and $\bar{r}.$  In order to find $\phi$ and $b$, we start from
(\ref{dhat}) and write

\be
dt=\frac{1}{\phi}\left\{d\hat{t}-\phi bd\bar{r}\right\},
\label{3.1}
\ee
and so

\be
dt^2=\frac{1}{\phi^2}\left\{d\hat{t}^2-2b\phi d\hat{t}d\bar{r}+b^2\phi^2d\bar{r}^2\right\}.
\label{3.2}
\ee 
Using these in (\ref{frw-1}) we obtain

\begin{eqnarray}
ds^2&=&\frac{1}{\phi^2}\left(\frac{\dot{R}^2}{R^2}\bar{r}^2-1\right)d\hat{t}^2+
\frac{2}{\phi}\left\{-b\left(-1+\frac{\dot{R}^2}{R^2}\bar{r}^2\right)-H\bar{r}\right\}d\hat{t}d\bar{r}+
\nonumber\\
&&\ \ \ \
\ \ \ \ \
\ \ \ \ \
\ \ \ \left\{1+b\left[b\left(-1+\frac{\dot{R}^2}{R^2}\bar{r}^2\right)+\frac{2\dot{R}}{R}\bar{r}\right]\right\}d\bar{r}^2.
\label{3.3}
\end{eqnarray}
The outline of our procedure for matching the components of the metric (\ref{3.3}) to the components of the metric
({\ref{EF}) is as follows:  We first set the coefficient of
$d\bar{r}^2$ equal to zero in (\ref{3.3}) and solve for the function $b.$  We then match the coefficients
of the cross terms $d\hat{t}d\bar{r}$
by setting the $d\hat{t}d\bar{r}$ coefficient in (\ref{3.3}) equal to $-2$ and then solve for the value of
$\phi$ on the shock surface.  Using the values of $b$ and $\phi$ in the matching of the $d\hat{t}^2$ coefficients, we
obtain a formula for the shock surface.  Finally, we show that the shock surface is non-characteristic for the PDE
(\ref{exact}), implying that we can use the initial condition for $\phi$ to solve for $\phi$ in a neighborhood of the
shock surface, and thereby determine a nonsingular coordinate $\bar{t}=\bar{t}(t,r).$ 

To begin, set the coefficient of $d\bar{r}^2$ equal to zero in (\ref{3.3}) to obtain

\be
1+b\left\{b\left(-1+H^2\bar{r}^2\right)+2H\bar{r}\right\}=0.
\label{3.4a}
\ee
This simplifies to 

\be
\left(H\bar{r}b+1\right)^2=b^2,
\nonumber
\ee 
so that

\be
b=\frac{-1}{H\bar{r}^+_-1}.
\label{3.4}
\ee
In order to ensure that $b,$ given in (\ref{3.4}), is nonsingular and of a single sign, we choose the plus sign if
$\dot{R}>0,$ (the case of a White Hole explosion), and the minus sign if $\dot{R}<0,$ (the case of collapse to a Black
Hole).  Matching the cross terms in (\ref{EF}) and (\ref{3.3}) yeilds

\be
\frac{2}{\phi}\left\{-b\left(-1+H^2\bar{r}^2\right)-H\bar{r}\right\}=-2.
\label{3.5}
\ee    
Using (\ref{3.4a}), we obtain

\be
\phi=H\bar{r}+\frac{1}{b}=^-_+1
\label{3.6}
\ee
at the shock surface.  Here, consistent with (\ref{3.4}), we take $+1$ if $H=\frac{\dot{R}}{R}>0,$ and $-1$ if
$H=\frac{\dot{R}}{R}<0$.  Matching the $d\hat{t}^2$ terms in (\ref{EF}) and (\ref{3.3}) gives

\be
-A=\frac{1}{\phi^2}\left(H^2\bar{r}^2-1\right).
\label{3.7}
\ee    
Since $k=0$ and $p=0,$ the FRW equations (\ref{frw1}), (\ref{frw2}) imply

\be
H^2=\frac{\kappa}{3}\rho.
\label{3.8}
\ee 
Using (\ref{3.8}) and (\ref{A}) together with $\phi^2=1$ in (\ref{3.7}) yields the equation for the shock surface, 

\be
M=\frac{\kappa}{6}\rho(t) \bar{r}^3.
\label{3.9}
\ee 
Solving (\ref{3.9}) for $\bar{r}$ as a function of $t$ gives the shock surface
$\bar{r}=\bar{r}(t)$ in the FRW coordinate.  It follows from the FRW equations (\ref{frw1}), (\ref{frw2}) in the case $p=0,$
that
$R(t)^3\rho(t)$ is constant on FRW solutions, \cite{wein}.  Using this in (\ref{3.9}) implies that $r=r_0=const.$
describes the interface in the OS solution.  For example, if we choose $R(t_0)=1$ at present time $t=t_0,$ then

\be
r_0=\left\{\frac{6M}{\kappa \rho(t_0)}\right\}^{1/3}\label{shockconst}
\ee
is constant on the interface, and thus $\bar{r}(t)=R(t)r_0$ describes the motion of the interface in the OS solution.  The
fact that
$r=const.$ along the interface implies that the interface moves with the galaxies of the $p=0$ FRW metric, and hence neither
mass nor momentum crosses the interface. 
  From this we can also see that the shock surface is
non-characteristic for the PDE (\ref{exact}).  Indeed, the characteristic curves for (\ref{exact}) satisfy 

$$
\frac{d\bar{r}}{dt}=-\frac{1}{b}=(\dot{R}r_{0})^+_-1,
$$ 
on the shock surface, where the $^+_-$ agrees with the sign of $\dot{R}.$   Thus we see that
$\frac{d\bar{r}}{dt}=(\dot{R}r_0)^+_-1$ along the characteristic, and $\frac{d\bar{r}}{dt}=\dot{R}r_0$
along the shock surface, so the two can never be equal.  We conclude that the shock
surface is never characteristic for equation (\ref{exact}), and hence $\phi,$ as well as the coordinate $\hat{t},$ can be
defined in a neighborhood of the interface. 

Note that the shock surface equation (\ref{3.9}) tells us that $\bar{r}$ goes from zero to infinity as
the FRW density goes from infinity to zero.  For example in the case $\dot{R}>0,$ this tells us that the FRW universe starts
at the White Hole singularity of the EF metric at the instant $t=0$ of the Big Bang.  From this time onward, the shock
surface $\bar{r}(t)=\left(\frac{6M}{\kappa\rho(t)}\right)^{1/3}$ continues out until the
the Hubble distance $H^{-1}=\left(\frac{3}{\kappa\rho(t)}\right)^{1/2},$ (c.f. (\ref{3.8})), catches up to the shock surface on the FRW side of
the shock, which happens at the FRW time
$t_s$ when
$\bar{r}(t_s)=2M,$ i.e., when the shock surface lies on the event
horizon
of the outer EF metric.  As the FRW time increases from
$t_s$ to
$\infty,$ the interface continues on out to infinity, staying inside the Hubble length on the FRW side of the shock, and
outside the Black Hole on the EF side. 

Note also that the coordinate $\bar{r}$ is a spacelike coordinate that measures arc-length distance in the FRW metric at
fixed time $t$ in the FRW coordinate system $(t,\bar{r})$ on the FRW side of the shock, but on the EF side of the shock, the
coordinate
$\bar{r}$ is lightlike in
$EF$ coordinates
$(\hat{t},\bar{r}),$ and changes from timelike to spacelike in standard Schwarzschild coordinates $(\bar{r},\bar{t})$ as
the shock surface passes through the event horizon of the EF metric---even so, the coordinate identification
$(t,r)\rightarrow(\hat{t},\bar{r})$ is regular in a neighborhood of the shock surface for all shock positions
$0<\bar{r}<\infty.$  This is no contradiction, because the coordinate vector field
$\frac{\partial}{\partial\bar{r}}$ remains undetermined until a choice of complementary coordinate is specified.

In conclusion, we have that the interface between the FRW and Schwarzschild metrics in the $k=0$ OS solution
is a contact discontinuity that traverses a geodesic of the Schwarzschild metric.  The total mass of
the FRW metric behind the shock wave interface is finite and constant in time, and emerges from the White Hole
singularity of the ambient Schwarzschild metric at $\bar{r}=0,$ the instant of the Big Bang. After a finite proper time, the solution
continues out through the event horizon at Schwarzschild time
$t=-\infty,$  and continues to expand forever into the asymptotically flat Schwarzshild spacetime outside the Black
Hole,  where
$\frac{2M}{\bar{r}}<1.$  In this regime the solution agrees with the $k=0$ OS solution given in
\cite{smolte5}\footnotemark[8].  
\footnotetext[8]{It is well known that the spacetime coordinatized by either the EF or Schwarzschild
coordinates is not a geodesically complete spacetime. That is, from the point of view of the (complete) Kruskal development of the
Schwarzschild spacetime, the EF metric covers only two of the four quadrants determined by
the event horizon in the Kruskal diagram; namely, the standard quadrant outside the Black Hole together with the quadrant
which contains the Black Hole singularity--or else under time reversal, the standard quadrant outside the Black Hole
together with the region containing the White Hole singularity.  On the other hand, the OS solution is a geodesically complete spacetime because, from the
Kruskal point of view, it consists of the Schwarzschild solution on one side, and the FRW solution on the other side, of the geodesic defined by the
trajectory of the interface of discontinuity.  Thus, by removing a neighborhood of either the White Hole or the Black Hole event horizon, the OS matching also
eliminates the incomplete geodesics that emerge from $t=-\infty,$ respectively $t=+\infty,$, depending on the time orientation of the solution,
\cite{misnthwh}.}     

\section{The TOV Solution Inside the Black Hole}\label{sect4}
\setcounter{equation}{0}

The standard Tolman-Oppenheimer-Volkoff (TOV) metric models a static fluid sphere in general relativity.  It was proved in
\cite{smolte3} that the standard TOV metric for a static fluid sphere, by itself, does not admit Black Holes, and can
only exist when
$\frac{2M}{\bar{r}}<1.$ In  this section we derive the equations for the analogue of the TOV solution {\it inside the  Black Hole}.

The TOV solutions {\it inside the  Black Hole} are used in Sections
\ref{sect5} and
\ref{sect6} to extend the OS solution of Section \ref{sect3} to the case of non-zero pressure, inside the Black Hole.  When
$p\neq0,$ energy and momentum cross the interface, and so the contact discontinuity of the OS solution must be replaced by a shock wave discontinuity. Thus
in  order to extend our shock matching techniques beyond the Hubble length when $p\neq0,$ we must replace the outer
Schwarzschild metric of the OS solution with a metric that contains matter, and satisfies $2M/\bar{r}>1.$ 
The TOV metric {\it inside the  Black Hole} is the simplest metric that satisfies these conditions.  We refer to it as
 TOV because the components depend only on the
radial coordinate
$\bar{r},$ as in the standard TOV metric, but now
$\bar{r}$ is timelike.  (As in the Schwarzschild metric, the roles of space and time are
interchanged inside the Black Hole.)  In shock matching with FRW metrics, this new TOV metric {\it inside the  Black Hole} will play the role of a
transitional solution that mediates the mass flux across the shock interface during the time after the Big Bang when the densities are large, and up until the
time when the solution has settled down to a zero pressure OS expansion.

The usual ansatz for a TOV metric takes the form

\be
ds^2=-Bd\bar{t}^2+A^{-1}d\bar{r}^2+\bar{r}^2d\Omega^2,
\label{4.1}
\ee
where $A(\bar{r})$ and $B(\bar{r})$ depend only on the coordinate $\bar{r}.$  Here 

\be
A(\bar{r})=1-\frac{2M(\bar{r})}{\bar{r}},
\label{4.2}
\ee
and when $\bar{r}>2M,$ $M(\bar{r})$ has the interpretation as the total mass inside the ball of radius $\bar{r}.$ 
Thus our assumption that the TOV metric lies {\it inside the Black Hole} or {\it inside the Schwarzschild radius} is
equivalent to the assumption that 
$$
A<0.
$$
We now obtain the Einstein equations for a perfect fluid under the assumption that the fluid is
co-moving with respect to a metric of form (\ref{4.1}), assuming that $A$ and $B$ depend only on $\bar{r},$
but now assuming that
$A<0.$ The stress tensor for a perfect fluid takes the form,

\be
T_{ij}=(\bar{\rho}+\bar{p})\bar{w}_i\bar{w}_j+\bar{p}\bar{g}_{ij},
\label{4.5}
\ee
and since $\bar{r}$ is the
timelike coordinate for the TOV metric in $(\bar{r},\bar{t})$-coordinates when $A<0,$ the assumption that the fluid
is co-moving with the TOV metric
{\it inside the  Black Hole} implies that, c.f. \cite{wein},
\be
(\bar{w}_{0},\bar{w}_{1},\bar{w}_{2},\bar{w}_{3})=(\bar{w}_{\bar{r}},\bar{w}_{\bar{t}},\bar{w}_{\theta},\bar{w}_{\phi})=\left(\frac{1}{\sqrt{-A}},0,0,0\right).
\label{4.6}
\ee
In this case we obtain

\be
T_{00}\equiv
T_{\bar{r}\bar{r}}=(\bar{\rho}+\bar{p})\left(\frac{1}{-A}\right)+\bar{p}\left(\frac{1}{A}\right)=-\frac{\bar{\rho}}{A},
\label{4.7}
\ee
and

\be
T_{11}\equiv
T_{\bar{t}\bar{t}}=\bar{p}\bar{g}_{11}=-\bar{p}B.
\label{4.8}
\ee
Using MAPLE we find, 
 
\be
G_{00}\equiv G_{\bar{r}\bar{r}}=-\frac{\bar{r}AB'-B+AB}{\bar{r}AB},
\label{4.3}
\ee

\be
G_{11}\equiv G_{\bar{t}\bar{t}}=-\frac{B}{\bar{r}^2}\left\{\bar{r}A'-1+A\right\}.
\label{4.4}
\ee
From (\ref{4.7})-(\ref{4.4}), the Einstein equations $G_{00}=\kappa T_{00}$ and $G_{11}=\kappa T_{11}$ reduce to 

\be
\frac{B'}{B}=\frac{1}{\bar{r}}\left(\frac{1-A}{A}\right)+\kappa\frac{\bar{\rho}}{A},
\label{4.9}
\ee
and

\be
A'=\frac{1-A}{\bar{r}}+\kappa \bar{p}\bar{r},
\label{4.10}
\ee
respectively.\footnotemark[9]  
\footnotetext[9]{Beware that MAPLE's convention is that
$-G_{MAPLE}=G,$ so the Einstein equations are $-G_{MAPLE}=\kappa T.$} 
From \cite{wein}, equation (5.45), we also have

\be
-\bar{p}'=(\bar{p}+\bar{\rho})\frac{d}{d\bar{r}}\ln\left\{\frac{1}{\sqrt{-A}}\right\},
\label{4.11}
\ee
which simplifies to

\be
\bar{p}'=\frac{\bar{p}+\bar{\rho}}{2}\frac{A'}{A}=\frac{\bar{p}+\bar{\rho}}{2A}\left\{\frac{1-A}{\bar{r}}+\kappa
\bar{p}\bar{r}\right\}.
\label{4.12}
\ee
Using (\ref{4.2}), we obtain the system

\begin{eqnarray}
\bar{p}'&=&\frac{\bar{p}+\bar{\rho}}{2}\frac{A'}{A}=\frac{\bar{p}+\bar{\rho}}{2A}\left\{\frac{1-A}{\bar{r}}+\kappa
\bar{p}\bar{r}\right\},\label{4.13}\\ A'&=&\frac{1-A}{\bar{r}}+\kappa \bar{p}\bar{r},\label{4.14}\\
\frac{B'}{B}&=&\frac{1}{\bar{r}}\left(\frac{1-A}{A}\right)+\kappa\frac{\bar{\rho}}{A}\label{4.15}.
\end{eqnarray}
Alternatively, using the unknown $N=1-A$ instead of $A,$ (a variable convenient for our subsequent analysis), we obtain the
equivalent system,

\begin{eqnarray}
\bar{p}'&=&\frac{\bar{p}+\bar{\rho}}{2}\frac{N'}{N-1},\label{4.16}\\ N'&=&-\left\{\frac{N}{\bar{r}}+\kappa
\bar{p}\bar{r}\right\},\label{4.17}\\
\frac{B'}{B}&=&-\frac{1}{N-1}\left\{\frac{N}{\bar{r}}+\kappa\bar{\rho}\right\}.\label{4.18}
\end{eqnarray}

Note that the essential reason why the equations for the TOV
metric {\it inside the  Black Hole}\  take a different form than the standard TOV equations \lq\lq outside the Black
Hole\rq\rq, is that the assumption of {\it co-moving} puts the nonzero component of ${\bf w}$ on the
timelike coordinate $\bar{r}$ when $A<0,$ and on the timelike coordinate $\bar{t}$ when $A>0.$    We conclude
that system (\ref{4.16})-(\ref{4.18}) defines a new class of gravitational metrics which describe spacetimes that evolve
{\it inside the Black Hole}.  

The TOV equations
{\it inside the  Black Hole} describe a time dependent metric in which the metric components, together with the fluid variables, are constant at
each fixed value of the TOV timelike coordinate
$\bar{r}.$  Like the FRW metric, the TOV metric {\it inside the  Black Hole} describes a fluid with
pressure that emerges from a White Hole singularity at
$\bar{r}=0,$ the instant of the Big Bang, with one important difference: unlike the FRW metric, for the TOV metric {\it inside the Black Hole}, the {\em total
mass} is {\em constant} on each spacelike slice
$\bar{r}=constant.$  In the next section we match the TOV metric {\it inside the Black Hole} to $k=0$ FRW metrics across shock interfaces in order to make
the expanding FRW universe {\em finite}, in the sense that the total mass on each spacelike hypersurface is finite.   

\section{Shock Matching Inside the Black Hole}\label{sect5}
\setcounter{equation}{0}

In this section we derive the equations that describe the matching of a general $k=0$ FRW metric to a TOV
metric {\it inside the  Black Hole}, at a shock wave interface across which energy and momentum are conserved.  The equations, given in system
(\ref{ueqn})-(\ref{Neqn}) below, describe the simultaneous time evolution of the shock position together with the TOV metric {\it inside the Black Hole}
such that the resulting metric matches a given $k=0$ FRW metric Lipschitz continuously across the shock, and such that conservation of energy and momentum
hold at the shock.  The conservation constraint, given in (\ref{cc}) below, determines the TOV pressure, and this is used to close the equations, c.f
\cite{smolte2}. Equations (\ref{ueqn})-(\ref{Neqn}) then determine all other unknowns in the TOV metric beyond the shock, and guarantee conservation, once
the FRW metric is assigned.  In particular, the equations guarantee that there are no delta function sources produced by the (second order) Einstein
equations due to the lower order smoothness (Lipschitz continuity) of the metrics at the shock, c.f. (i)-(iv) below. The success of the method relies on the
fact that once Lipschitz continuity is imposed, the single conservation condition (\ref{cc}) alone guarantees the two conservation constraints
of the Rankine-Hugoniot jump conditions at the shock, c.f. \cite{smolte1}, and Theorem \ref{theorem3} below. 

Solutions of equations (\ref{ueqn})-(\ref{Neqn}) are formally time-reversible without the imposition of an entropy condition. In the next section we
formulate an entropy condition that agrees with the entropy condition of gas dynamics, and we use this to construct exact
solutions of (\ref{ueqn})-(\ref{Neqn}) that describe a class of time-irreversible entropy satisfying shocks in which the TOV density and pressure satisfy the
physical bounds
$0<\bar{p}<\bar{\rho}.$  

To model the expanding universe with a spherical shock wave at FRW position
$\bar{r}=\bar{r}(t),$ we assume that the FRW metric lies in the region $\bar{r}<\bar{r}(t),$ and the TOV metric lies beyond the shock
wave at
$\bar{r}>\bar{r}(t).$  The corresponding matching for a standard TOV metric outside the Black Hole, (and inside the Hubble
length), was accomplished in
\cite{smolte2,smolte3,smolte9}.  In this section we do not use an EF type regularization of the TOV metric, (c.f.
(\ref{sschwarz})), but rather we employ standard Schwarzschild coordinates, avoiding the singularity at $A=0$ by working either
inside or outside the Black Hole separately. 

Our proceedure for shock matching is as follows:  We first identify the shock surface across which a $k=0$ FRW metric matches
a TOV metric {\it inside the Black Hole}, such that the matching is Lipschitz continuous across the interface, and such that we have a smooth matching of
the spheres of symmetry, (c.f., \cite{smolte1}, Lemma 9, equation (5.3)).   The Lipschitz matching of the metrics is guaranteed by a
non-characteristic condition that always holds outside the Black Hole, when $A>0,$ c.f. (\ref{shockdot3}) below.  Given this matching, we then determine a
conservation constraint that guarantees that the Rankine-Hugoniot jump conditions

\be
[T^{\mu\nu}]n_{\mu}=0,
\label{consconst0}
\ee 
hold across the shock. (Here, as usual, $[\cdot]$ denotes the jump in a quantity across the shock interface, $n_{\mu}$ are
the covariant components of the normal vector ${\bf n}$ to the shock surface, and assuming spherical symmetry, we need only
require (\ref{consconst0}) for $\nu=0,1.$)  We can then apply \cite{smolte1}, Lemma 9, which states that for metrics matched Lipschitz continuously across
the shock, with a smooth matching of the spheres of symmetry, the Rankine-Hugoniot jump conditions
(\ref{consconst0}) imply that the following equivalencies are also valid at the shock:
\vspace{.2cm} 

{\bf (i)}
The extrinsic curvature is continuous across the shock,

{\bf (ii)} The Riemann and Einstein curvature tensors, viewed as second order operators on
the metric components, produce no delta function sources on the shock,

{\bf (iii)} In a neighborhood of each point on the shock surface there exist coordinate transformations
such  that in the new coordinates, all second derivatives of the metric components are bounded a.e., and

{\bf (iv)} At each point on the shock surface there exist coordinate frames that are locally Lorentzian.
\vspace{.2cm}

In the papers \cite{smolte1,smolte2,smolte5,smolte9}, the main idea for carrying out this procedure in the case $A>0,$ was to show that the
single condition
\be
[T^{\mu\nu}]n_{\mu}n_{\nu}=0,
\label{consconst}
\ee 
alone implies the jump conditions (\ref{consconst0}), and hence {\bf (i)}-{\bf (iv)}.  However, we must modify this idea in the case $A<0$ because it turns
out that the condition (\ref{consconst}) has a non-physical solution which everywhere violates the non-characteristic condition used to guarantee the
Lipschitz matching of the metrics at the shock.  For this reason, to obtain the conservation condition when the shock wave lies inside the Black Hole,
we verify the Rankine-Hugoniot conditions directly.

To start, we match the FRW to the TOV metric when $k=0$ and $A<0.$ The $k=0$ FRW metric in the usual $(t,r)$ coordinates is given by,

\begin{equation}
\label{frw0}
ds^2=-dt^2+R(t)^{2}\left\{dr^2+r^2d\Omega^2\right\}.
\end{equation}
In (\ref{frwschwarz1})-(\ref{B-1}) we showed that the mapping $(t,r)\rightarrow(\bar{r},\bar{t})$ that takes metric
(\ref{frw0}) to standard Schwarzschild coordinates in which the metric takes the form,

\begin{equation}
\label{frwschwarz-1}
ds^2=-B_{FRW}d\bar{t}^2+
A_{FRW}^{-1}d\bar{r}^2+\bar{r}^2d\Omega^2,
\end{equation}
is given by

\be
\bar{r}&=&R(t)r,\label{B-2a}
\ee 
and 

\be
d\bar{t}=(\psi C)dt-(\psi E)d\bar{r},
\label{exact-1a}
\ee
where   

\be
C&=&1-H^2\bar{r}^2,\label{C1}\\
E&=&-\bar{r}H.\label{E1}
\ee
(Recall that $\frac{\partial}{\partial\bar{r}}$ is timelike in the coordinate system $(\bar{r},\bar{t})$ and spacelike in the coordinate system $(t,\bar{r})$
when $A_{FRW}<0,$ $H^{-1}>1.$) Here $\psi$ can be taken to be any solution of 
\begin{equation}
\label{exact-2}
\frac{\partial}{\partial \bar{r}}(\psi C)+\frac{\partial}{\partial t}(\psi E)=0,
\end{equation}
 determined from initial data on any non-characteristic surface, (c.f. the paragraph following (\ref{char})).  Under this
coordinate transformation, we obtain

\be
A_{FRW}^{-1}&=&\frac{R^2}{R^2-\bar{r}^2\dot{R}^2}=\frac{1}{1-\left(\bar{r}H\right)^2},\label{A-2}\\
B_{FRW}&=&\frac{1}{\psi^2}\frac{R^2}{R^2-\bar{r}^2\dot{R}^2}=\frac{1}{\psi^2}\frac{1}{1-\left(\bar{r}H\right)^2},\label{B-2}
\ee
c.f. (\ref{frwschwarz1})-(\ref{B-1}).
We use the subscript $FRW$ to
distinguish the FRW metric coefficients $A_{FRW}= 1-\frac{2M_{FRW}}{\bar{r}}$
and $B_{FRW}$ from the TOV metric coefficients
$A=1-\frac{2M}{\bar{r}}$ and
$B,$ that appear in the TOV line element

\begin{equation}
\label{TOV-1}
ds^2=-Bd\bar{t}^2+
A^{-1}d\bar{r}^2+\bar{r}^2d\Omega^2,
\end{equation}
in standard Schwarzschild coordinates.  By shock matching, using (\ref{frw1}) and (\ref{A-2}), we find that
\begin{equation}
\label{Mfrw}
M_{FRW}(t,\bar{r})=\frac{1}{2}\bar{r}^3H^2=\frac{\kappa}{6}\rho(t)\bar{r}^3.
\end{equation}

Assume now that the
$(\bar{r},\bar{t})$ coordinates that describe the TOV and FRW metrics in standard Schwarzschild coordinates, actually
represent a single coordinate system for the pair of metrics matched across a shock surface where the metrics agree; that is,
where
$A=A_{FRW}$ and $B=B_{FRW}.$  Setting $A_{FRW}=A$ and using (\ref{Mfrw})
gives 

\be
\frac{2M_{FRW}}{\bar{r}}=(H\bar{r})^2=\frac{\kappa}{3}\rho\bar{r}^2=\frac{2M}{\bar{r}},
\label{shockdot4}
\ee
from which we deduce the following conservation of mass condition that must hold at the shock surface, and which is
independent of
$\psi:$
\be
\frac{\kappa}{6}\rho(t)\bar{r}^3=M(\bar{r}).
\label{5.3}
\ee
Equation (\ref{5.3}) implicitly defines the shock surface
$\bar{r}=\bar{r}(t),$ which then determines the position $r=r(t)=\frac{\bar{r}(t)}{R(t)}$ of the shock in the original FRW
coordinates
$(t,r).$  By (\ref{shockdot4}) we also see that 

\be
\label{interpretN}
N\equiv 1-A=\frac{2M}{\bar{r}}=(\bar{r}H)^2,
\ee
and so 

\be
\label{shockdistance}
\bar{r}=\sqrt{N}H^{-1}
\ee 
holds on the shock surface.  Using (\ref{shockdistance}), we can interpret $N$ from the TOV metric as follows:  if
$(t,\bar{r})$ is a point on the shock surface, then
$\sqrt{N(t,\bar{r})}$ is equal to the number of Hubble lengths from the center of the FRW spacetime to the shock surface as
measured at fixed time
$t$ on the FRW side of the shock.  Since $N>1$ if and only if $A<0,$ it follows that the shock wave lies beyond one
Hubble length from the FRW center when $A<0.$ Note that $N$ is a convenient variable because it appears in system (\ref{4.16})-(\ref{4.18}).

 Setting $B_{FRW}=B$ at the shock surface and using (\ref{B-2}), we obtain the initial condition for
the integrating factor $\psi$ of the FRW metric on the shock surface (\ref{5.3}); namely,

\be
\psi=^+_-\left\{(1-H^2\bar{r}^2)B(\bar{r})\right\}^{-1/2}=^+_-\left\{AB\right\}^{-1/2},
\label{5.4}
\ee 
where we have used that $1-H^2\bar{r}^2=A<0$ at the shock.   Note that $AB>0,$ and the choice of sign in (\ref{5.4})
determines the time orientation for $\bar{t},$ c.f. (\ref{exact-1a}).

 We now derive a condition that guarantees
that the surface $\bar{r}=\bar{r}(t)$ is non-characteristic when
$A<0.$\footnotemark[10]\footnotetext[10]{This is no moot point.  Indeed, in previous work, (c.f. \cite{smolte1,smolte2}),
the authors used (\ref{consconst}) to obtain the conservation constraint across the shock surface when $A>0,$ but we must be careful here because a
calculation shows that one of the solutions of (\ref{consconst}) is everywhere characteristic for (\ref{exact-2}) when $A<0.$  It follows that
(\ref{consconst0}) does not hold across the characteristic surface, and so there is a solution of (\ref{consconst}) that does not represent an actual weak
solution of
$G=\kappa T$ when $A<0.$  It is for this reason that we go directly to (\ref{consconst0}) to construct the conservation constraint
in the case $A<0$ below.}   To this end, differentiate (\ref{5.3}) with respect to FRW time
$t$ to obtain,
 
\be
M'\dot{\bar{r}}=\frac{\kappa}{6}\dot{\rho}\bar{r}^3+\frac{\kappa}{2}\rho\bar{r}^2\dot{\bar{r}}.
\label{shockdot}
\ee
Using (\ref{frw2}) and (\ref{4.17}) in (\ref{shockdot}) we obtain the following formula for the speed 
$\dot{\bar{r}},$ the speed at which the shock wave is receding from an observer fixed at the FRW origin:

\be
\dot{\bar{r}}=\frac{\rho+p}{\rho+\bar{p}}H\bar{r}.
\label{shockdot1}
\ee
For future reference we record that this directly implies

\be
\dot{r}&=&\frac{H\bar{r}}{R}\left(\frac{p-\bar{p}}{\rho+\bar{p}}\right). \label{rdot}
\ee
On the other hand, in light of (\ref{char}), the
characteristics for (\ref{exact-2}) satisfy

\be
\dot{\bar{r}}=-\left\{\bar{r}H\right\}^{-1}+\bar{r}H.
\label{shockdot2}
\ee
Thus the shock surface is characteristic if and only if the two speeds in (\ref{shockdot1}) and (\ref{shockdot2}) are equal,
which holds if and only if 

\be
1-\left\{\frac{\rho+p}{\rho+\bar{p}}\right\}=\frac{1}{(H\bar{r})^2}.
\label{shockdot3}
\ee
Using (\ref{shockdot4}) in (\ref{shockdot3}) we conclude that the shock surface is
non-characteristic for the PDE (\ref{exact-2}) if and only if

\be
A\neq\frac{\rho+p}{p-\bar{p}}.
\label{noncharcondt}      
\ee
The inequality (\ref{noncharcondt}) is not in general ruled out on a timelike shock surface when $A<0.$ However,
(\ref{noncharcondt}) immediately implies the following lemma which we record for use in Section 6:

\begin{Lemma}
\label{charcond}
If $\rho,p,\bar{\rho},\bar{p}$ are all positive on the shock surface $\bar{r}=\bar{r}(t)$ defined implicitly by (\ref{5.3}),
then the surface is non-characteristic for the PDE (\ref{exact-2}) so long as $p>\bar{p}$ and $A<0.$
\end{Lemma}

Assuming that (\ref{noncharcondt}) holds, the above procedure defines a
mapping
$(t,r)\rightarrow(\bar{r},\bar{t}),$ defined in a neighborhood of any point on the shock surface $(\bar{r}(t),t),$ such
that, under this coordinate identification, the FRW metric matches the TOV metric
Lipshitz continuously across the shock surface.  Note also that as in the OS solution, the coordinate
$\bar{r}$ is timelike  in the Schwarzschild coordinates $(\bar{r},\bar{t})$ when $A<0,$ but spacelike in the FRW coordinates
$(t,\bar{r}),$ where the coordinate $\bar{r}$ measures arclength distance in each spatial slice $t=const.$

It remains to analyze
the Rankine-Hugoniot jump conditions (\ref{consconst0}).  The following theorem gives a formulation of the conservation
constraint that is amenable to analysis.

\begin{Theorem}\label{Theoremcc}\label{theorem3}
Assume that the coordinate mapping $(t,r)\rightarrow(\bar{r},\bar{t})$ defines a Lipshitz continuous matching of an FRW
metric to a TOV metric {\it inside the Black Hole}, in a neighborhood of a point $P$ across a
non-characteristic shock surface
$\bar{r}=\bar{r}(t)$ defined by (\ref{5.3}), so that that $N>1$ at $P.$ Assume further that $p,\rho>0$ at $P.$  Then the
Rankine-Hugoniot jump relations (\ref{consconst0}) together with the equivalent conditions (i)-(iv) listed
after (\ref{consconst0})) all hold at $P$, if and only if the following single conservation constraint holds at the point $P:$ 
\be
\label{cc}
\bar{p}=\frac{p-\left(\frac{\bar{\rho}+p}{\rho-\bar{\rho}}\right)\frac{1}{N}\rho}
{1+\left(\frac{\bar{\rho}+p}{\rho-\bar{\rho}}\right)\frac{1}{N}}.
\ee
\end{Theorem}

Note that (\ref{cc}) immediately implies that if $\rho>\bar{\rho},$ then also $p>\bar{p},$ consistent with an outgoing
explosion of an inner FRW metric into an outer TOV metric.  (The construction of an explicit example of such a shock wave
is the topic of the next section.)  Solving for
$\bar{\rho}$, it follows that condition (\ref{cc}) is equivalent to the condition

\be
\label{cc1}
\bar{\rho}=\frac{-\left(\rho+\bar{p}\right)p+(p-
\bar{p})N\rho}{(\rho+\bar{p})+(p-\bar{p})N}.
\ee
Note also that from (\ref{cc1}) we see that if $p=\bar{p}=0,$ then also $\bar{\rho}=0,$
the conditions of the OS solution.

Before giving the proof of Theorem \ref{Theoremcc}, we first use the conservation constraint (\ref{cc1}) to derive the
equations that describe the time evolution of the shock surface and TOV metric {\it inside the Black Hole} that matches a
given $k=0$ FRW metric at a shock wave positioned beyond one Hubble length $N>1.$  To start, assume
that $\rho(t),p(t)$ and $R(t)$ are known functions of the FRW time
$t$  that determine a unique equation of state $p(\rho),$ and let $\sigma(t)=p(t)/\rho(t).$   The matching condition
(\ref{5.3}) written in the form

\be
\label{matchingN}
\rho=\frac{3N}{\kappa\bar{r}^2},
\ee
gives $\rho$ as a function of $N$ and $\bar{r}$ at the shock,  and substituting this into the conservation constraint
(\ref{cc1}) and using $p=p(\rho)$ gives $\bar{\rho}$ as a function of $\bar{p},N$ and $\bar{r}.$   The first two equations
(\ref{4.16}), (\ref{4.17}) of the TOV system (\ref{4.16})-(\ref{4.18}) then close upon substituting the resulting expression
$\bar{\rho}=\bar{\rho}(\bar{p},N,\bar{r})$ for $\bar{\rho}$ in the first equation (\ref{4.16}).  The resulting system, 

\be
\bar{p}'&=&\frac{\bar{p}+\bar{\rho}(\bar{p},N,\bar{r})}{2}\frac{N'}{N-1},\label{2by2-1}\\
N'&=&-\left\{\frac{N}{\bar{r}}+\kappa\bar{p}\bar{r}\right\},\label{2by2-2}
\ee
then forms a non-autonomous system of two equations in the unknowns $(\bar{p},N)$ as a function of the independent
variable
$\bar{r},$ the shock position.  (Again, ``prime'' denotes $\frac{d}{d\bar{r}}.$) Solving (\ref{2by2-1}), (\ref{2by2-2}) gives
$\bar{p}(\bar{r})$ and
$N(\bar{r})$ subject to the initial conditions
 
\be
\label{icondts}
\bar{p}=\bar{p}_0,\ N=N_0,\ \ at\ \ \bar{r}=\bar{r}_0.
\ee
Assuming the FRW solution $\rho(t),$ $p(t),$ and $R(t)$ is given, we can obtain $\bar{r}$ as a function of $t$ from (\ref{matchingN}), whereby we conclude
that
$\bar{p}(\bar{r})$ and
$N(\bar{r})$ determine the entire shock wave solution.  For example, to set the position of the shock at present time in the FRW metric to be at
$\bar{r}=\bar{r}_0,$ choose
$\rho_0=present\ density,$ define

\be
\label{Nzero}
N_0=\frac{\kappa}{3}\rho_0\bar{r}_0^2,
\ee
(c.f. use (\ref{matchingN})).  This leaves $\bar{p}_0$ as a free parameter.  Once we know the solution $(\bar{p}(\bar{r}),N(\bar{r})),$ we can
use (\ref{matchingN}) to obtain the FRW density $\rho$ as a function of $\bar{r},$ and inverting the relation

\be
\label{rhoofbarr-1}
\rho(t)=\frac{3N(\bar{r})}{\kappa\bar{r}^2},
\ee
gives the shock position $\bar{r}=\bar{r}(t),$ the distance from the shock wave to the FRW center at FRW time $t.$  The
FRW shock position is then

\be
\nonumber
r(t)=\frac{\bar{r}(t)}{R(t)}.
\ee
The only other unknown in the problem is the metric coefficient $B(\bar{r})$ from the TOV metric, which we get by integrating
the equation (\ref{4.18})\footnotemark[11]\footnotetext[11]{Note that (\ref{Bformula}) completes the definition of the TOV
metric, and so the integrating factor $\psi$ for the shock matching is determined by (\ref{5.4}), and from this the non-characteristic condition
(\ref{noncharcondt}) implies the Lipschitz matching of the metrics.},

\be
\label{Bformula}
B(\bar{r})=B_0\exp{\left\{-\int_{r_0}^r\frac{1}{N(\xi)-1}\left(\frac{N(\xi)}{\xi}+\kappa\bar{\rho}(\xi)\right)d\xi\right\}}.
\ee
Note that for an outgoing shock wave in an expanding FRW metric, it follows from (\ref{2by2-2}) that $N$ decreases and $\bar{r}$ increases in forward FRW
time, and so $N\rightarrow\infty,$ $\bar{r}\rightarrow0$ would correspond to the Big Bang from the FRW point of view. 

We conclude that the problem of constructing FRW-TOV shock
waves with
$N>1$ reduces to the analysis of the
$2\times2$ system (\ref{2by2-1}), (\ref{2by2-2}).  The next theorem shows that when we make the
change of variable $\bar{p}\rightarrow u=\frac{\bar{p}}{\rho}$ and take $N$
to be the independent variable, the resulting
equivalent system has the nice property that the equations are coupled to the FRW metric only through the variable $\sigma=\frac{p}{\rho}.$ 

\begin{Theorem}
\label{TheoremuN}
Under the change of variables
\be
u=\frac{\bar{p}}{\rho},\ v=\frac{\bar{\rho}}{\rho},\ \sigma=\frac{p}{\rho}, \label{uvdef}
\ee
system (\ref{2by2-1}), (\ref{2by2-2}) with conservation constraint (\ref{cc1}) transforms to the equivalent system 

\be
\frac{du}{dN}=-\left\{\frac{(1+u)}{2(1+3u)N}\right\}\left\{\frac{(3u-1)(\sigma-u)N+6u(1+u)}{(\sigma-u)N+(1+u)}\right\},\label{ueqn}
\ee
\be
\frac{d\bar{r}}{dN}&=&-\frac{1}{1+3u}\frac{\bar{r}}{N},\label{Neqn}
\ee
with conservation constraint
\be
\label{cc-3}
v=\frac{-\sigma\left(1+u\right)+(\sigma-u)N}{(1+u)+(\sigma-u)N}.
\ee
For such solutions, the speed of the shock interface relative to the fluid comoving on the FRW side of the shock, is given by
\be
s=R\dot{r}=\sqrt{N}\left(\frac{\sigma-u}{1+u}\right). \label{rdot-0}
\ee
\end{Theorem}

Note that the dependence of (\ref{ueqn})-(\ref{cc-3}) on
the FRW metric is only through the variable $\sigma,$ and hence (\ref{ueqn}) is coupled to (\ref{Neqn}) only through the function $\sigma.$  In particular,
if we assume that
$\sigma$ is constant in the FRW metric, (a reasonable model problem, c.f. \cite{smolte2}), then equation (\ref{ueqn})
uncouples from equation  (\ref{Neqn}) to form a non-autonomous scalar equation in $u$
and $N$ which is amenable to phase plane analysis.  This is the starting point for the exact solutions constructed in the
next section.  

Note too that since (\ref{Neqn}) implies that $\frac{dN}{d\bar{r}}<0,$ and $\sqrt{N}$ gives the number of Hubble lengths to the shock wave,
c.f. (\ref{shockdistance}), equation (\ref{Neqn}) implies that the number of Hubble lengths from the FRW center to the shock wave {\em decreases} in time,
as claimed in the introduction. 

Since system system (\ref{ueqn}), (\ref{Neqn}) is equivalent to system (\ref{2by2-1}}), (\ref{2by2-2}), the former system also determines all quantities in
the shock wave solution, but now as a function of the variable
$N.$  For example, a solution of system (\ref{ueqn}), (\ref{Neqn}) is
determined by the initial conditions
$u=u_0$ and
$\bar{r}=\bar{r}_0$ at
$N=N_0,$ which entails assigning three constants.  Alternatively, if we ask that the shock wave be positioned $\sqrt{N_0}$
Hubble lengths from the FRW center at the present FRW value of the density $\rho=\rho_0,$ then by (\ref{matchingN}), the
initial shock position is determined by

\be
\label{exact-1}
\bar{r}_0=\sqrt{\frac{3N_0}{\kappa\rho_0}},
\ee
and there is then a one parameter family of such solutions $(u(N),\bar{r}(N))$ determined by the remaining free parameter

\be
\label{exact-3}
u_0=\frac{\bar{p}_0}{\rho_0}.
\ee
Once we know the solution $(u(N),\bar{r}(N)),$ we can use

\be
\label{exact-4}
\rho=\frac{3N}{\kappa\left[\bar{r}(N)\right]^2},
\ee
to obtain $\rho(N),$ the FRW density $\rho$ as a function of $N$ at the shock.  Since the FRW metric is known, knowing
$\rho(N)$ determines
$p(N)$ and $\sigma(N)=\frac{p(N)}{\rho(N)},$ and using these in (\ref{cc-3}) we obtain $v(N)$ and
$\bar{\rho}(N)=v(N)\rho(N).$ Finally, the TOV metric coefficient $B$ is determined as a function of $N$
by integrating the equation (\ref{4.18}) with respect to $N,$ yielding the formula

\be
\label{exact-5}
B(\bar{r})=B_0\exp{\left\{-\int_{N_0}^N\frac{1}{\xi-1}\left(\frac{\xi}{\bar{r}(\xi)}+\kappa\bar{\rho}(\xi)\right)d\xi\right\}}.
\ee
To connect $N$ with the FRW time $t,$ invert the relation (\ref{exact-4}), 
obtaining $N(t).$  Since $R(t)$ is assumed known, we can use

\be
\label{exact-7}
r(t)=\frac{\bar{r}(N(t))}{R(t)},
\ee
to get the FRW shock position $r$ as a function of $t.$  Relations (\ref{exact-1})-(\ref{exact-7}) formally determine an
exact shock wave solution of the Einstein equations for any given $(k=0)$ FRW metric, and any given solution of system
(\ref{ueqn}), (\ref{Neqn}).

\vspace{.2cm}

\noindent{\bf Proof of Theorem \ref{TheoremuN}:}  Equation (\ref{cc-3}) follows directly from (\ref{cc1}) upon dividing
through by
$\rho$ and making the substitutions in (\ref{uvdef}).  Equation (\ref{Neqn}) follows directly from (\ref{4.17}) upon noting
that $\kappa\bar{p}\bar{r}=\frac{3uN}{\bar{r}},$ a consequence of (\ref{rhoofbarr-1}).  From (\ref{Neqn}) we have

\be
\label{uprime-0}
\frac{1}{\bar{r}}=-\frac{1}{(1+3u)}\frac{N'}{N}.
\ee
To verify
(\ref{ueqn}), start with

\be
\label{uprime-1}
u'=\left(\frac{\bar{p}}{\rho}\right)'=\frac{\bar{p}'}{\rho}-\frac{\bar{p}}{\rho^2}\rho'.
\ee
Then by (\ref{2by2-1}) and (\ref{cc-3}),

\be
\label{uprime-2}
\frac{\bar{p}'}{\rho}=\frac{u+v}{2}\frac{N'}{N-1}=
\frac{1}{2}\left\{\frac{(1+u)(\sigma-u)}{(1+u)+(\sigma-u)N}\right\}N',
\ee
and using (\ref{matchingN}),

\be
\label{uprime-3}
\rho'=\left\{\frac{3N}{\kappa\rho\bar{r}^2}\right\}'=\rho\left(\frac{N'}{N}-\frac{2}{\bar{r}}\right)=
\rho\left(\frac{3(1+u)}{1+3u}\right)\frac{N'}{N},
\ee
where we have used (\ref{uprime-0}).  Thus,

\be
\label{uprime-4}
\frac{\bar{p}}{\rho^2}\rho'=\left(\frac{3u(1+u)}{1+3u}\right)\frac{N'}{N},
\ee
and so using (\ref{uprime-2}) and (\ref{uprime-4}) in (\ref{uprime-1}) gives

\be
\label{uprime-5}
u'&=&\frac{1}{2}\left\{\frac{(1+u)(\sigma-u)}{(1+u)+(\sigma-u)N}\right\}N'-\left(\frac{3u(1+u)}{1+3u}\right)\frac{N'}{N}
\nonumber\\
&=&-\frac{(\sigma-u)(3u^2+2u-1)N+2u(1+u)^2}{2(1+3u)[(1+u)+(\sigma-u)N]N}N',\nonumber\\
&=&-\left\{\frac{(1+u)N'}{2(1+3u)N}\right\}\left\{\frac{(3u-1)(\sigma-u)N+6u(1+u)}{(\sigma-u)N+(1+u)}\right\},\nonumber
\ee
which upon dividing by $N'$ verifies (\ref{ueqn}).  Finally, since the fluid on the FRW side of the shock is assumed to be co-moving with the radial
coordinate $r$ of the FRW coordinate system, it follows that $s=R\dot{r}$ gives the speed of the shock relative to the FRW fluid,
(c.f., the discussion after Theorem 4 of \cite{smolte2}), and so (\ref{rdot-0}) follows directly from (\ref{rdot}). $\Box$
\vspace{.2cm}

\noindent{\bf Proof of Theorem \ref{Theoremcc}:}  Assume that we are given an FRW metric and TOV metric {\it inside the Black
Hole} that match Lipshitz continuously across a smooth, non-characteristic shock surface $\bar{r}=\bar{r}(t)$ defined in a
neighborhood of point $P,$ such that the hypotheses of Theorem \ref{theorem3}  hold.  We show that (\ref{consconst0}) holds if
and only if (\ref{cc}) holds.  We use the following lemma:

\begin{Lemma}
\label{LemmajumpT}
On the shock surface, in $(\bar{r},\bar{t})$ coordinates, we have,

\be
\label{jumpinT}
\left[T\right]^{\mu\nu}&\equiv&\left[T_{FRW}^{\mu\nu}-T_{TOV}^{\mu\nu}\right]\\
&=&
\left[\begin{array}{cc}
(\rho+p)N+(\bar{\rho}+p)(1-N)\ \ \ \ \ \ \ \ \psi\sqrt{N}(\rho+p)\\ \ \ \ \ \ \psi\sqrt{N}(\rho+p)\ \
\ \ \ \ \ \psi^2\left\{(\rho+\bar{p})+(p-\bar{p})N\right\}
\end{array}\right],\nonumber
\ee
and

\be
\label{shocknormal0}
\bar{n}_0&=&\psi\left(N\frac{p-\bar{p}}{\rho+\bar{p}}+1\right),\\
\label{shocknormal1}
\bar{n}_1&=&-\sqrt{N}\left(\frac{\rho+p}{\rho+\bar{p}}\right),
\ee
where $\bar{n}_{\mu}d\bar{x}^{\mu}=\bar{n}_0d\bar{r}+\bar{n}_1d\bar{t}$ is the covariant normal to the shock surface.

\end{Lemma}
We use Lemma \ref{LemmajumpT} to verify Theorem \ref{Theoremcc}, and postpone the proof of Lemma \ref{LemmajumpT} until the
end. To verify that (\ref{consconst0}) is equivalent to (\ref{cc}) at a point $P$ on the shock surface, it suffices to show that
$det\left([T]^{\mu\nu}\right)=0$ holds at $P$ if and only if (\ref{cc}) holds at $P,$
and that when this holds, the shock normal
${\bf n}$ is in the kernel of $[T].$  Indeed, if (\ref{cc}) holds at $P,$ and we know ${\bf n}$ is in the kernel of
$[T]^{\mu\nu},$ then (\ref{consconst0}) and (\ref{consconst}) hold at $P,$ and so our general theory in \cite{smolte1} would imply that
(i)-(iv) hold at $P$ as well.  Conversely, if (i)-(iv) and (\ref{consconst0}) all hold at $P,$ then the RH jump conditions (\ref{consconst0}) alone imply that
$det\left([T]^{\mu\nu}\right)=0$ holds at $P,$ and hence we would have (\ref{cc}) at $P$ as well. 

We first show that $det\left([T]^{\mu\nu}\right)=0$ holds at $P$ if and only if (\ref{cc}) holds at $P.$  But by (\ref{jumpinT}),

\be
\label{detT-1}
&det\left([T]^{\mu\nu}\right)=\left\{(\rho+p)N+(\bar{\rho}+p)(1-N)\right\}_I
\left\{\psi^2[(\rho+\bar{p})+(p-\bar{p})N]\right\}&\nonumber\\
&\ \ \ \ \ \ \ \ \ \ \ \ \ \ \ \ \ \ \ -\psi^2N(\rho+p)^2=0,&  
\ee
and solving for $\bar{p}$ gives

\be
\label{detT-2}
\bar{p}&=&\left[-\frac{N(\rho+p)^2}{\left\{(\rho-\bar{\rho})N+\bar{\rho}+p\right\}_I}+(\rho+pN)\right]\frac{1}{N-1}
\nonumber\\
&=&\frac{\left\{(\rho+pN)\left\{\cdot\right\}_I-N(\rho+p)^2\right\}}{(N-1)\left\{\cdot\right\}_I}\nonumber\\
&=&\frac{(N-1)\left\{p(\rho-\bar{\rho})N-\rho(\bar{\rho}+p)\right\}}{(N-1)\left\{(\bar{\rho}+p)+(\rho-\bar{\rho})N\right\}},\nonumber 
\ee
from which (\ref{cc}) follows at once.  (Note that we have assumed without loss of generality that $\{\cdot\}_I\neq0,$ which
is valid because, if $\{\cdot\}_I=0,$ then (\ref{detT-1}) implies $\rho=p=0,$ and $\rho=0$ violates
$N>1,$ e.g.,  
$\dot{R}\neq0,$ c.f. (\ref{detT-1}), (\ref{interpretN}).)  Thus it remains only to show that if $det\left([T]^{\mu\nu}\right)=0$ holds at $P,$ then the shock
normal
${\bf n}$ at $P$ is in the kernel of $[T].$  To this end, note that by (\ref{shocknormal0}) and (\ref{shocknormal1}),

\be
\label{ndotT}
\bar{n}_{\mu}[T]^{\mu1}&=&\psi^2\left\{\left(N\frac{p-\bar{p}}{\rho+\bar{p}}+1\right)\sqrt{N}(\rho+p)\right.\\\nonumber
&&\left.-\sqrt{N}\left(\frac{\rho+p}{\rho+\bar{p}}\right)\left((\rho+\bar{p})+(p-\bar{p}\right)N\right\}=0,
\ee
and so $\bar{n}_{\mu}[T]^{\mu0}=0$ as well because we assume $det[T]=0,$ and $[T]$ is a $2\times2$ matrix 
\footnotemark[12]\footnotetext[12]{Although (\ref{ndotT}) appears to be a miracle here, this is to be expected from \cite{smolte1,smolte10}, where in the case
$A>0,$ the single condition
$[T^{\mu\nu}]n_{\mu}n_{\nu}=0$ alone implied the equivalencies (i)-(iv), as well as the Rankine-Hugoniot conditions (\ref{consconst0}).  One also expects
this in light of the fact that the Einstein equations for a spherically symmetric metric in standard Schwarzschild coordinates, contains only one second
order, (i.e. for Lipschitz metrics, one weak), equation, so we would expect only one jump condition.}.   The proof of Theorem
\ref{Theoremcc} is complete, once we give the
\vspace{.2cm}

\noindent{\bf Proof of Lemma \ref{LemmajumpT}:}  To verify (\ref{jumpinT}), we calculate $T_{TOV}^{\mu\nu}$ and
$T_{FRW}^{\mu\nu},$ the stress tensors in barred coordinates ${\bf \bar{x}}=(\bar{r},\bar{t}),$ at the shock.  (Recall
that $\bar{x}^0=\bar{r}$ is the timelike coordinate, and $\bar{x}^1=\bar{t}$ is spacelike because $N>1.$)  Using the formula 

\be
\label{perfectfluidstress}
T_{TOV}^{\mu\nu}=(\bar{\rho}+\bar{p})\bar{u}^{\mu}_{TOV}\bar{u}^{\nu}_{TOV}+\bar{p}g^{\mu\nu},
\ee
and the assumption that the TOV fluid is co-moving in barred coordinates,

\be
\label{perfectfluidstress-1}
\bar{u}^{\mu}=\left[\begin{array}{c}
\sqrt{-A}\\0
\end{array}\right]^{\mu}=\left[\begin{array}{c}
\sqrt{N-1}\\0
\end{array}\right]^{\mu},\nonumber
\ee
we obtain

\be
\label{perfectfluidstress-2}
T_{TOV}^{\mu\nu}=(\bar{\rho}+\bar{p})\left[\begin{array}{cc}
-A\ \ \ 0\\\ \ 0\ \ \ \ 0
\end{array}\right]^{\mu\nu}+\bar{p}A\left[\begin{array}{cc}
1\ \ \ \ \ 0\\0\ -\psi^2
\end{array}\right]^{\mu\nu}.\nonumber
\ee
Similarly,  
\be
\label{perfectfluidstressfrw}
T_{FRW}^{\mu\nu}=(\rho+p)\bar{u}^{\mu}_{FRW}\bar{u}^{\nu}_{FRW}+pg^{\mu\nu},
\ee  
but in this case the assumption that the FRW fluid is co-moving in FRW unbarred coordinates $x^{\alpha}=(t,r)^{\alpha},$
implies that

\be
\label{utransform}
\bar{u}^{\mu}=\frac{\partial \bar{x}^{\mu}}{\partial x^{\alpha}}u^{\alpha},
\ee
where

\be
\label{ucomponents}
u^{\alpha}=\left[\begin{array}{c}
1\\0
\end{array}\right]^{\alpha},
\ee
gives the components of the FRW fluid velocity in standard FRW coordinates $(t,r).$
\vspace{.2cm}

\noindent{\bf Claim:}  When $k=0,$ we have

\be
\label{claim}
\frac{\partial \bar{x}^{\mu}}{\partial x^{\alpha}}=\left[\begin{array}{cc}
\sqrt{N}\ \ \ \ R\\\ \ \ \ \psi\ \ \ \ \psi R\sqrt{N}
\end{array}\right]^{\mu}_{\alpha}.
\ee
\vspace{.2cm}

\noindent{\bf Proof of Claim:}  The coordinate mapping $\bar{r}=Rr$ implies

\be
\nonumber
\frac{\partial\bar{x}^0}{\partial t}&=&\dot{R}r=H\bar{r},\\
\frac{\partial\bar{x}^0}{\partial r}&=&R,
\ee
which verifies the the first row of (\ref{claim}) upon using (\ref{interpretN}), $H\bar{r}=\sqrt{N}.$  Moreover, by
(\ref{exact-1a})-(\ref{E1}), 
\be
d\bar{t}=\psi (1-H^2\bar{r}^2)dt+\psi H\bar{r}d\bar{r}.
\label{exact-11}
\ee
Substituting 

$$d\bar{r}=d(Rr)=\dot{R}rdt+Rdr=H\bar{r}dt+Rdr$$
into (\ref{exact-11}) gives

\be
d\bar{t}=\psi\left(dt+H\bar{r}Rdr\right),
\label{exact-111}
\ee
from which we can deduce the second row of (\ref{claim}), which verifies the Claim. 

Continuing now from (\ref{utransform}):

\be
\nonumber
\bar{u}^{\mu}=\left[\begin{array}{cc}
\sqrt{N}\ \ \ \ R\\\ \ \psi\ \ \ \ \psi R\sqrt{N}
\end{array}\right]^{\mu}_{\alpha}\left[\begin{array}{c}
1\\0
\end{array}\right]^{\alpha}=\left[\begin{array}{c}
\sqrt{N}\\\psi
\end{array}\right]^{\mu}.
\ee
Using this in (\ref{perfectfluidstressfrw}) gives

\be
\label{perfectfluidstressfrw-1}
T_{FRW}^{\mu\nu}=(\rho+p)\left[\begin{array}{cc}
\ \ \ N\ \ \ \ \ \psi\sqrt{N}\\\ \psi\sqrt{N}\ \ \ \ \ \ \psi^2
\end{array}\right]^{\mu\nu}+pA\left[\begin{array}{cc}
1\ \ \ \ \ 0\\\ \ 0\ \ -\psi^2
\end{array}\right]^{\mu\nu}.
\ee
Using (\ref{perfectfluidstressfrw-1}) together with (\ref{perfectfluidstress-2}) gives

\be
\nonumber
[T]^{\mu\nu}&=&[T_{FRW}^{\mu\nu}-T_{TOV}^{\mu\nu}]\\
&=&
\left[\begin{array}{cc}
(\rho+p)N+(\bar{\rho}+p)(1-N)\ \ \ \ \ \ \ \ \psi\sqrt{N}(\rho+p)\\ \ \ \ \ \ \psi\sqrt{N}(\rho+p)\ \
\ \ \ \ \ \psi^2\left\{(\rho+\bar{p})+(p-\bar{p})N\right\}
\end{array}\right],\nonumber
\ee
which is (\ref{jumpinT}).  

Finally, we use the Claim to verify (\ref{shocknormal0}) and (\ref{shocknormal1}) as follows.  Let

\be
\phi(t,r)\equiv r-r(t)=0,
\label{5.34}
\ee
where $r(t)$ denotes the shock surface in FRW $(t,r)$-coordinates.  Then

\be
d\phi=-\dot{r}dt+dr=n_0dt+n_1dr,
\nonumber
\ee
gives a covariant normal $n_{\alpha}dx^\alpha$ to the shock surface in  ${\bf x}=(t,r)$ coordinates,
\be
n_0&=&-\dot{r},\nonumber\\
n_1&=&1.\label{5.36}
\ee  
In
${\bf\bar{x}}=(\bar{r},\bar{t})$-coordinates,

\be
\bar{n}_{\mu}=\frac{\partial x^{\alpha}}{\partial \bar{x}^{\mu}}n_{\alpha},\label{5.36a}
\ee
where $\frac{\partial x^{\alpha}}{\partial \bar{x}^{\mu}}$ is the inverse of the matrix (\ref{claim}),

\be
\frac{\partial x^{\alpha}}{\partial \bar{x}^{\mu}}=\frac{1}{det\left|\frac{\partial \bar{x}^{\mu}}{\partial
x^{\alpha}}\right|}\left[\begin{array}{cc}
\psi R\sqrt{N}\ \ \ \ -R\\\ \ -\psi\ \ \ \ \sqrt{N}
\end{array}\right]^{\alpha}_{\mu}.
\label{5.36b}
\ee
Using (\ref{5.36}) and (\ref{5.36b}) in (\ref{5.36a}) and neglecting the irrelevant factor $\frac{1}{det\left|\frac{\partial \bar{x}^{\mu}}{\partial
x^{\alpha}}\right|},$ we obtain, (to within a minus sign),

\be
\bar{n}_{\mu}=\left(\begin{array}{cc}
\psi \left[\dot{r}R\sqrt{N}+1\right],-\dot{r}R-\sqrt{N}
\end{array}\right)_{\mu},
\label{5.36c}
\ee
and substituting (\ref{rdot}) in the form

\be
\nonumber
\dot{r}=\frac{\sqrt{N}}{R}\frac{p-\bar{p}}{\rho+\bar{p}},
\ee
into (\ref{5.36c}), then gives the formulas for $\bar{n}_{\mu}$ given in (\ref{shocknormal0}), (\ref{shocknormal1}).
The proof of Lemma \ref{LemmajumpT} is complete.  $\Box$
\vspace{.2cm}

We summarize the results in this section as follows:  System (\ref{ueqn})-(\ref{Neqn}) describes the TOV metrics that
match a given $k=0$ FRW metric across a shock wave discontinuity when $A<0,$ assuming that the conservation constraint
(\ref{cc-3}) is satisfied.  Solutions of these equations, together with the initial conditions in (\ref{exact-1})-(\ref{exact-7}), determine weak
solutions of the Einstein equations containing a shock wave interface across which the metric is only Lipschitz continuous,
and such that conditions (i)-(iv) (after (\ref{consconst})) are satisfied.  Since the resulting solutions are formally
time-reversible, it still remains to impose a physically meaningful entropy condition.  The entropy condition for shocks
determines the time orientation of the solution, c.f.
\cite{smol}.  In the next section we construct a class of exact solutions of these equations which satisfy an entropy
condition that agrees with the entropy condition of gas dynamics in the non-relativistic limit.


\section{Exact Shock Wave Solutions When $A<0.$}\label{sect6}
\setcounter{equation}{0}

In this section we derive a class of exact solutions of equations (\ref{ueqn})-(\ref{Neqn}) in the case when the FRW pressure
is given by the equation of state

\be
p=\sigma\rho,
\label{7.1}
\ee
where $\sigma$ is assumed to be constant, 

\be
\label{sigma}
0<\sigma<1.
\ee
In this section, as an entropy condition, we impose the condition that the shock be compressive, a condition sufficient
to choose the physically relevant stable shocks in classical gas dynamics, \cite{glim,lax,smol}. That is, we will require that the pressure and density be
larger on the side of the shock that receives the mass flux.  Exact solutions satisfying the equation of state
(\ref{7.1}) were constructed in
\cite{smolte2} for the case
$A>0,$ so we can interpret the results here as an extension of the results in \cite{smolte2} to the case $A<0.$  However, in 
\cite{smolte2}, the conservation constraint led to a TOV equation of state also of the form (\ref{7.1}), but in our case here,
the TOV pressure is not so simple due to the fact that the conservation constraint (\ref{cc-3}) is of a different form. 

Assuming (\ref{7.1}), the solution to the FRW equations (\ref{frw1}), (\ref{frw2}) is given in Theorem \ref{theorem2},
equations (\ref{frw3-1})-(\ref{inftyexact}), and we assume
an expanding universe, ($\dot{R}>0$), with initial conditions
$R=0$ at $t=0,$ and $R=1,$ at $t=t_0.$  The solution has one assignable constant which can be taken to be $t_0,$ $\rho_0$ or $H_0,$ since

\be
\rho_0&=&\frac{4}{3\kappa(1+\sigma)^2}\frac{1}{t_0^2},\label{initial-1}
\ee
\be
H_0=\left[\frac{3(1+\sigma)}{2}\right]\frac{1}{t_0},\label{initial-2}
\ee
where we interpret $t_0$ as \lq\lq present time\rq\rq in the FRW metric\footnotemark[13],
\footnotetext[13]{Since we are assuming the idealized equation of state (\ref{7.1}) for the FRW metric, the value of the Hubble constant at present time
alone determines the solution.  In the standard model of cosmology based on a $k=0$ FRW metric, \cite{blaugu}, the solution is determined by two assignable
constants, the Hubble constant
$H_0$ and the background radiation temperature $T_0,$ \cite{blaugu,smolte9}.  Assuming (\ref{7.1}), the freedom to assign $T_0$ is represented by the freedom
to assign
$\sigma.$  Also, in the standard model the galaxies follow the particle paths of the matter field after the time when the pressure of the matter
field is essentially zero.  In the model (\ref{7.1}), the matter is co-moving with respect to the
radial coordinate
$r,$ and hence the particle paths follow
$\dot{r}=0.$  Loosely speaking, we refer to the motion along the particle paths as the motion {\it of the galaxies}
in the cosmological interpretation of a general FRW metric. Thus we say that
{\it the shock wave is exploding outward through the galaxies} when the shock wave satisfies $\dot{r}>0,$} and the relations (\ref{exact-1})-(\ref{exact-7})
hold.

We now solve the TOV shock equations (\ref{ueqn})-(\ref{Neqn}), to obtain a
shock wave solution that matches this FRW metric for all
$R(t)$ in the interval between $R=0$ and $R=1.$  Since  (\ref{ueqn}) is coupled to
(\ref{Neqn}) only through the function $\sigma,$ our assumption that
$\sigma$ is constant in the FRW metric implies that solutions of system (\ref{ueqn})-(\ref{Neqn}) are determined by
solutions of the scalar non-autonomous equation (\ref{ueqn}).  Making the change of variable $S=1/N,$ 
(we do this to transform the ``Big Bang''
$N\rightarrow \infty$ over to rest point at $S\rightarrow0,$ c.f. \cite{smolte9}), equation (\ref{ueqn}) takes the
form

\be
\frac{du}{dS}=\left\{\frac{(1+u)}{2(1+3u)S}\right\}\left\{\frac{(3u-1)(\sigma-u)+6u(1+u)S}{(\sigma-u)+(1+u)S}\right\},\label{-6-1}\label{12}
\ee
where $u=\bar{p}/\rho.$  We now construct solutions of (\ref{-6-1}) that model the ``Big
Bang'' as a localized explosion with an outgoing blast wave emanating from $\bar{r}=0$ at time $t=0.$  Thus, motivated by classical gas
dynamics,  we seek solutions of (\ref{-6-1}) that satisfy the entropy conditions,

\be
&0<\bar{p}<p,&\label{entropy1}\\&0<\bar{\rho}<\rho,&
\label{entropy2}
\ee
and meet the physical bounds on the TOV equation of state
\be
&0<\bar{p}<\bar{\rho}.&\label{state1}
\ee   
Conditions (\ref{entropy1}), (\ref{entropy2}) for outgoing shock waves imply that the shock wave is compressive. The condition
(\ref{state1}) implies that the TOV equation of state is physically reasonable.  Note that the conditions $N>1$ and $0<\bar{p}<p$ restrict the domain of
(\ref{-6-1}) to the region
$0<u<\sigma<1,$ $0<S<1.$   The following theorem shows that the inequalities (\ref{entropy1}), (\ref{entropy2}) and
(\ref{state1}) can all be expressed in terms of
$u$ and $S$ alone, and are all implied by a single inequality.

\begin{Theorem}
\label{entropyinuandS}
Assume that 
\be
\label{6.1}
0<u<\sigma,
\ee

\be
\label{6.2}
0<S<1,
\ee
and the conservation constraint (\ref{cc-3}) holds. Then the bounds (\ref{entropy1})-(\ref{state1}) are all implied by the single condition

\be
\label{6.5}
S<\left(\frac{1-u}{1+u}\right)\left(\frac{\sigma-u}{\sigma+u}\right)\equiv E(u).
\ee
\end{Theorem}

\noindent{\bf Proof of Theorem \ref{entropyinuandS}:}  Multiplying (\ref{6.1}) through by $\rho>0$ gives (\ref{entropy1}). 
To verify that $\bar{\rho}<\rho,$ write (\ref{cc-3}) in the form

\be
\label{cc-4}
v=\frac{-\sigma S+a}{S+a},
\ee
where
\be
\label{cc-4a}
a=\frac{\sigma-u}{1+u}>0,
\ee
and observe that $\bar{\rho}<\rho$ is equivalent to $v<1,$ and by (\ref{cc-4}), $v<1$ if and only if $-\sigma<1.$    
To verify (\ref{state1}), observe that $\bar{p}<\bar{\rho}$ is equivalent to $\frac{v}{u}>1,$ and by
(\ref{cc-4}), this is equivalent to 
$$
\frac{-\sigma S+a}{(S+a)u}>1,
$$
which is easily seen to be equivalent to (\ref{6.5}).  Thus it remains only to verify $0<\bar{\rho}.$ But observe that $\bar{\rho}>0$ is equivalent to $v>0,$
and by (\ref{cc-4}), this is equivalent to
\be
\label{6.4-1}
S<\frac{1}{\sigma}\frac{\sigma-u}{1+u},  
\ee
which is implied by (\ref{6.5}) because
\be
E(u)<\frac{1}{\sigma}\frac{\sigma-u}{1+u}.
\nonumber
\ee
The proof of the theorem is complete. $\Box$

\subsection{Analysis of the equations}

The purpose of this section is to prove the following theorem:

\begin{Theorem}\label{theorem6}
For every $\sigma,$ $0<\sigma<1,$ there exists a unique solution $u_{\sigma}(S)$ of (\ref{-6-1}), such that (\ref{6.1}) and (\ref{6.5}) hold on the
solution for all $S,$ $0<S<1,$  and on this solution,
\be
0<u_{\sigma}(S)<\bar{u},\label{usigmabound}
\ee 
\be
&\lim_{S\rightarrow0}u_{\sigma}(S)=\bar{u},&\label{Sarrow0}
\ee
where
\be
\bar{u}=Min\left\{\sigma,1/3\right\},\label{baru}
\ee
and
\be
&\lim_{S\rightarrow1}\bar{p}=0=\lim_{S\rightarrow1}\bar{\rho}.&\label{Sarrow1}
\ee

\end{Theorem}

\noindent{\bf Proof of Theorem \ref{theorem6}:}  Solutions of (\ref{-6-1}) are determined by trajectories
of the autonomous system

\be
S'&=&2S(1+3u)\left\{(\sigma-u)+(1+u)S\right\}=F(S,u),\label{-6-3}\label{1}\\
u'&=&(1+u)\left\{-(1-3u)(\sigma-u)+6u(1+u)S\right\}=G(S,u),\nonumber\\\label{-6-2-1}\label{2}
\ee
which has two degenerate rest points, $S=0,$ $u=\sigma$ and $S=0,$ $u=1/3,$ in the $(S,u)$-phase plane.  Here
\lq\lq prime\rq\rq denotes differentiation with respect to a parameter
$\xi,$ and we recover equation (\ref{-6-1}) by eliminating $\xi$ via $du/dS=u'/S'.$  Since
$E(0)=1,$ (c.f. (\ref{6.5})), the ``initial condition''
$u_{\sigma}(1)=0$ is a consequence of the entropy inequality (\ref{6.5}), and thus the uniqueness of $u_{\sigma}(S)$ follows from uniqueness
of solutions for system (\ref{-6-3}), (\ref{-6-2-1}) because $u=0,$
$S=1,$ is a regular initial condition for the system.  Thus (\ref{Sarrow1}) follows from (\ref{uvdef}) and (\ref{cc-3}) because
$\lim_{S\rightarrow1}u_{\sigma}(S)=0.$   It remains to prove
the existence of the solution
$u_{\sigma}(S)$ on
$0<S<1,$ and to establish (\ref{Sarrow0}).

Our analysis of these equations is based on the construction of invariant regions.  Before we analyze the phase plane for system (\ref{-6-3}), (\ref{-6-2-1}),
first note that
$F(S,u)>0$ in the region
$0\leq u<\sigma,$
$0<S\leq1,$ and the isocline
$G=0$ is given by
\be
\label{Gisocline}
S=\frac{(\sigma-u)(1/3-u)}{2u(1+u)}\equiv h(u,\sigma).
\ee
Let $R_{\sigma}$ denote the region\footnotemark[14]\footnotetext[14]{We include the line $u=0$ in $R_{\sigma}$ in order that the initial point $u=0,$ $S=1$
for the solution trajectory lies in
$R_{\sigma}$.  However, the entropy conditions (\ref{entropy1}), (\ref{state1}) do not hold at the transitional
point $u=0,$
$S=0,$ the place where the solution $u_{\sigma}(S)$ continues naturally to the OS solution when $p<<1.$}
\be
R_{\sigma}=\left\{(S,u):0\leq u<\bar{u},\ \ 0< S\leq h(u,\sigma)\right\},\label{invR}
\ee 
c.f. Figure 1.

\begin{figure}\begin{center}
\includegraphics[scale=0.28,angle=270]{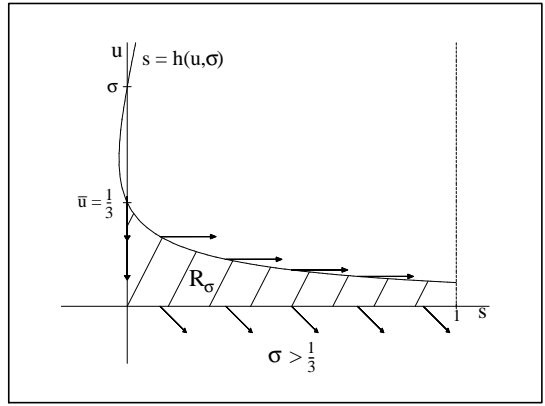}
\includegraphics[scale=0.28,angle=270]{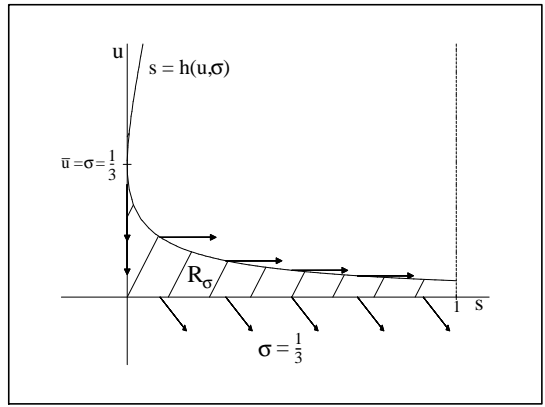}
\includegraphics[scale=0.28,angle=270]{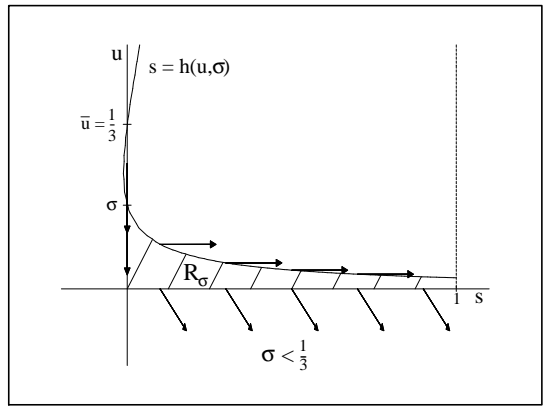}
\caption{The invariant region $R_\sigma$}
\end{center}\end{figure}

\begin{Lemma}\label{lemma3}
The region $R_{\sigma}$ is a negatively invariant region for system (\ref{-6-3}), (\ref{-6-2-1}) for each $\sigma$ between $0$ and $1.$ 
\end{Lemma}
\vspace{.2cm}

\noindent{\bf Proof:} To show that $R_{\sigma}$ is negatively invariant, it suffices to show that the vector field $(F,G)$ points out of or tangent to
$R_{\sigma}$ on the boundary of $R_{\sigma},$ as shown in Figure 1.  Now the vector field $(F,G)$ is tangent to the axis
$S=0$ because
$S=0$ is a solution trajectory of system (\ref{1}), (\ref{2}).  Since the boundary $u=0$ lies below the $G$-isocline, it follows that $G<0$ along the boundary
$u=0,$ and so the vector field $(F,G)$ points out of $R_{\sigma}$ along the lower boundary.  Finally,  the vector field points out of $R_{\sigma}$ along the upper boundary
$S=h$ because, for each  
$\sigma,$ $0<\sigma<1,$ $S=h$ is a $G$-isocline, $F$ is positive, and $S=h(u,\sigma)$ is a decreasing function of $u$ on $0\leq u<\bar{u}.$  Indeed,
$S=h(u,\sigma)$ is decreasing on $0\leq u<\bar{u}$ because, (using (\ref{Gisocline}), and prime for $\frac{d}{du}$), 

\be
\label{Gisocline-1}
h'=\frac{A'}{B}-A\left(\frac{B'}{B^2}\right)\leq0,
\ee
where $h=\frac{A}{B}$ and 
\be
A&=&(\sigma-u)(1/3-u),\nonumber\\
B&=&2u(1+u),\nonumber
\ee
and we use the fact that $A\geq0,$ $A'\leq0,$ $B\geq0,$ and $B'\geq0$ in $0\leq u<\bar{u}.$ 

Now the existence of the orbit $u_{\sigma}(S)$ on the entire interval $0<S\leq1$ is a consequence of the fact that $S=0,$ $u=\bar{u}$ is the only rest point
in the region $R_{\sigma}.$  That is, the unique trajectory of system (\ref{1}), (\ref{2}) starting from initial point $(S,u)=(1,0)\in R_{\sigma},$ must tend
to the unique rest point
$(0,\bar{u})$ in backward time $\xi\rightarrow-\infty.$  Moreover, since $G<0$ and $F>0$ in the interior of $R_{\sigma}$ it follows that $S$ and $u$ are
monotone along the orbit, so the orbit defines the trajectory $u=u_{\sigma}(S),$ $0<S\leq1,$  as well as its inverse, $S=S_{\sigma}(u),$ $0\leq u<\bar{u}.$ 

It remains only to verify that the entropy condition (\ref{6.5}) holds all along the solution, and for this it suffices to show that
\be
\Delta(u)\equiv E(u)-S_{\sigma}(u)\geq0, \label{Delta0}
\ee  
for all $u\in[0,\bar{u}).$  But
\be
\Delta(0)=0
\ee
because $E(0)=1=S_{\sigma}(0),$ and 
\be
\Delta(\bar{u})\geq0
\ee
because $E(\bar{u})\geq0$ and $S_{\sigma}(\bar{u})=0.$  Thus to verify that $\Delta(u)>0$ for $0\leq u<\bar{u},$ it suffices only to show that $\Delta(u)>0$
in a deleted neighborhood of $u=0,$ and that $\Delta'(u)=0$ has at most one root in $0\leq u<\bar{u}.$  But differentiating (\ref{6.5}) gives
\be
E'(u)=-2(1+\sigma)\frac{\sigma-u^2}{(\sigma+u)^2(1+u)^2}.\nonumber
\ee
Using this and (\ref{12}) at $u=0$ gives

\be
E'(0)=-2\frac{1+\sigma}{\sigma}=s'_{\sigma}(0),\nonumber
\ee
and
\be
E''(0)=4\frac{(1+\sigma)^2}{\sigma^2}>-6\frac{(1+\sigma)^2}{\sigma^2}=s''_{\sigma}(0),\nonumber
\ee 
which implies that $\Delta'(0)=0,$  $\Delta''(0)>0,$ so $\Delta(u)>0,$ near $u=0.$  
Moreover, by (\ref{6.5}) and (\ref{12}) we also see that $\Delta'(u)=0$ if and only if

\be
&&S^2\left\{(\sigma+u)^2(1+u)^3(1+3u)\right\}_a\label{quadraticE}\\
 &&\ \ \ \ \ +S\left\{(\sigma-u)(\sigma+u)^2(1+u)^2(1+3u)+6u(1+\sigma)(1+u)^2(\sigma-u^2)\right\}_b\nonumber\\
&&\ \ \ \ \ \ \ \ \ \ \ \ \ \ \ \ \ \ +\left\{(3u-1)(1+\sigma)(\sigma-u)(1+u)(\sigma-u^2)\right\}_c=0.\nonumber
\ee
But $\left\{\cdot\right\}_a>0,$ $\left\{\cdot\right\}_b>0,$ and $\left\{\cdot\right\}_c<0$ for $0<u<\bar{u}$ because
$\bar{u}=Min\left\{\sigma,1/3\right\},$ and thus the quadratic (\ref{quadraticE}) has exactly one
positive root.  We conclude that (\ref{Delta0}) holds.  Finally, to verify (\ref{Sarrow1}), we note that $u\rightarrow0$ as $S\rightarrow1,$ so 
$\bar{p}\rightarrow0.$  The fact that $\lim_{S\rightarrow1}\bar{\rho}=0$ follows directly from (\ref{cc-3}).   The proof
of Theorem
\ref{theorem6} is complete.  $\Box$
\vspace{.2cm}

Theorem \ref{theorem6} implies that the entropy conditions (\ref{entropy1}), (\ref{entropy2}) pick out the unique solution of system (\ref{-6-1}) emanating
from the initial point $S=1,$ $u\equiv\frac{\bar{p}}{\rho}=0=\bar{p}.$   
\vspace{.2cm}

\subsection{The shock speed}

In this section we estimate the shock speed $s_{\sigma}(S)\equiv s(u_{\sigma}(S)),$ the speed of the shock relative to an observer fixed relative to
the FRW fluid, along the solutions
$u_{\sigma}(S)$ of Theorem
\ref{theorem6},
$0<S<1,$ $0<u_{\sigma}(S)<\sigma.$  

\begin{Theorem}\label{theorem7}
Let $0<\sigma<1.$  Then the shock wave is everywhere subluminous, that is,
\be
\left|s_{\sigma}(S)\right|< 1, \label{ss1}
\ee
for all $0<S\leq1,$ if and only if 
\be
\sigma\leq 1/3. \label{ss2}
\ee
\end{Theorem}

\noindent{\bf Proof:}  Formula (\ref{rdot-0}) gives
\be
s_{\sigma}(S)=R\dot{r}=\frac{1}{\sqrt{S}}\left(\frac{\sigma-u}{1+u}\right), 
\label{shock}
\ee
and thus since $lim_{S\rightarrow0}u_{\sigma}(S)=\bar{u}=Min\{\sigma,1/3\},$ it follows at once that $lim_{S\rightarrow0}s_{\sigma}(S)=\infty$ if
$\sigma>1/3.$ Since $lim_{S\rightarrow0}u_{\sigma}(S)=\sigma$ when $\sigma\leq 1/3,$ we conclude from (\ref{shock}) that
the precise value of the shock speed $s_{\sigma}$ in the limit $S\rightarrow0$ depends on the asymptotic behavior of the solution $u_{\sigma}(S)$ as
$S\rightarrow0.$  This is addressed in the next subsection.

To verify (\ref{ss1}), (\ref{ss2}), we first prove the following lemma.

\begin{Lemma}\label{lemma4}
Assume that $0<\sigma\leq 1/3,$ and let $S=Q_a(u)\equiv Q(u)$ be defined by

\be
\label{Q}
Q_a(u)=a^2\frac{(\sigma-u)^2}{(1+u)^2}.  
\ee
Then the region 

\be {\cal Q}^a_{\sigma}\equiv\left\{(S,u):S\geq Q_a(u)\right\}\cap R_{\sigma},\label{invQ}
\ee
is a negatively invariant region for solutions of system (\ref{1}), (\ref{2}),
so long as

\be
a^2\leq \frac{1}{3\sigma}.
\label{abound1}
\ee
Moreover, for $\sigma=1/3$ and sufficiently small $\epsilon>0,$ there exists a $\delta>0,$ such that if 

\be
a^2\leq \frac{1}{1-\delta},
\label{abound2}
\ee
then solutions of (\ref{1}), (\ref{2}), starting in the region
\be {\cal Q}^{a}_{1/3}(\epsilon)\equiv {\cal Q}^{a}_{1/3}\cap\left\{(S,u):0\leq u< 1/3-\epsilon\right\} ,\label{invQ-1}
\ee
can only leave the region ${\cal Q}^a_{1/3}(\epsilon)$ through the boundary $u=1/3-\epsilon.$

\end{Lemma}
Note first that, assuming $\sigma\leq 1/3,$  the condition $a^2\leq \frac{1}{2\sigma}$ guarantees that $Q_a(u)<h(u)$ in the region $0\leq u<\sigma,$ $0<S\leq
1$ because by (\ref{Gisocline}), 
$Q_a(u)<h(u)$ is equivalent to
\be
a^2<\left(\frac{1/3-u}{\sigma-u}\right)\frac{(1+u)}{2u},\label{asquared}
\ee
and since
\be
\frac{1}{2\sigma}<\left(\frac{1/3-u}{\sigma-u}\right)\frac{(1+u)}{2u},\label{asquared1}
\ee
$a^2\leq\frac{1}{2\sigma}$ guarantees (\ref{asquared}) in $0\leq u<\sigma,$ $0<S\leq 1.$
It follows that when $a^2\leq\frac{1}{2\sigma},$ the region ${\cal Q}^a_{\sigma}$ is the region in the $(S,u)$-plane that lies above the curve $S=Q_a(u),$
above the curve $u=0,$ and below the isocline
$S=h(u).$ Moreover, when $a^2\leq\frac{1}{2\sigma},$ the initial point $u=0,$ $S=1$ for solution trajectory $u_{\sigma}(S)$ lies within the region ${\cal
Q}^a_{\sigma}$ because $Q_a(0)=a^2\sigma^2<1.$

Before proving the lemma, we first use Lemma \ref{lemma4} to complete the proof of Theorem \ref{theorem7} by verifying (\ref{ss1}).
Since the initial condition $S=1,$ $u=0$ lies within the 
region ${\cal Q}^a_{\sigma},$   we can estimate the shock speed all along the orbit $u_{\sigma}(S)$ by

\be
s_{\sigma}(u)=\frac{1}{\sqrt{S}}\left(\frac{\sigma-u}{1+u}\right)\leq
\frac{1}{a}\left(\frac{1+u}{\sigma-u}\right)\left(\frac{\sigma-u}{1+u}\right)=\frac{1}{a},
\ee
and so taking $a=\frac{1}{\sqrt{3\sigma}}<\frac{1}{\sqrt{2\sigma}},$ (optimal for (\ref{abound1})), we obtain

\be\label{shockspeeda}
s_{\sigma}(u)\leq \sqrt{3\sigma}<1,
\ee
for all $0<\sigma<1/3,$ $0\leq u\leq \sigma.$  This verifies (\ref{ss1}) for $0<\sigma<1/3.$  However, (\ref{shockspeeda}) does not provide a strict
inequality if $\sigma=1/3.$ 

To obtain the strict inequality (\ref{ss1}) in the case $\sigma=1/3,$ we
use that the initial condition
$S=1,$
$u=0$ also lies in
${\cal Q}^a_{1/3}(\epsilon)$ for $a=\frac{1}{\sqrt{1-\delta}},$ so if we assume
$u_{1/3}(S)\leq1/3-\epsilon,$ then (\ref{abound2}), (\ref{invQ-1}) imply that the shock speed in the region
${\cal Q}^a_{1/3}(\epsilon)$ can be estimated by

\be
s_{1/3}(u)\leq\frac{1}{a}= \sqrt{1-\delta}<1,
\ee
when $\sigma=1/3.$
Since $\epsilon$ is arbitrary, we conclude that (\ref{ss1}) holds for all $0<\sigma\leq 1/3,$ $0<S\leq1.$

The proof of Theorem \ref{theorem7} is complete, once we give the 
\vspace{.2cm}

\noindent{\bf Proof of Lemma \ref{lemma4}:}  Assume (\ref{ss2}).  To show that ${\cal Q}^a_{\sigma}$ is negatively invariant, it suffices to prove that the
vector field
$(F,G)$ restricted to the curve $S=Q_a(u),$ always points into the region below the curve.  That is, it suffices to prove that $|Q'(u)|\geq
\left|\frac{dS}{du}\right|$ on $S=Q_a(u)\equiv Q,$ where
$\frac{dS}{du}$ is given in (\ref{-6-1}).  Thus our condition is

\be
|Q'|=2a^2\frac{(1+\sigma)(\sigma-u)}{(1+u)^3}\geq \frac{2(1+3u)Q[(\sigma-u)+(1+u)Q]}{(1+u)|(3u-1)(\sigma-u)+6u(1+u)Q|}.\nonumber
\ee 
After some algebra, this reduces to
\be
\frac{(1+\sigma)}{(\sigma-u)(1+3u)}\geq\frac{(1+u)+(\sigma-u)a^2}{|(1-3u)(1+u)-6u(\sigma-u)a^2|}.\nonumber\\\label{invariantQ}
\ee
But the term inside the absolute value is positive so long as $a\leq\frac{1}{2\sigma}$ because then

\be
a^2\leq\frac{1}{2\sigma}\leq\left(\frac{1-3u}{\sigma-u}\right)\frac{1+u}{6u},\label{detsign}
\ee
as a result of (\ref{asquared1}).

Therefore, using that the term in the absolute value in (\ref{invariantQ}) is positive, we can solve (\ref{invariantQ}) for $a^2$ to obtain the following
condition equivalent to the condition that
$S\geq Q(u)$ is negatively invariant:
\be
a^2&\leq&\frac{(1+u)\left[(1+\sigma)(1-3u)-(1+3u)(\sigma-u)\right]}{(1+3u)(\sigma-u)^2+(1+\sigma)6u(\sigma-u)}.\nonumber\\
&=&\frac{(1+u)\left[(1+\sigma)\frac{(1-3u)}{(\sigma-u)}-(1+3u)\right]}{(1+3u)(\sigma-u)+6(1+\sigma)u}\label{equal1}
\ee
But using that 
$$\frac{1-3u}{\sigma-u}\geq3,$$
it follows that the inequality
\be
a^2&\leq&\frac{(1+u)[2+3(\sigma-u)]}{(1+3u)(\sigma-u)+6(1+\sigma)u}\equiv \phi_{\sigma}(u),\label{equal2}
\ee
guarantees (\ref{equal1}) all along the curve $S=Q_{a}(u).$  But $\phi_{\sigma}'(u)<0$ for $0\leq u\leq\sigma\leq1/3$ because
$\phi_{\sigma}(u)=\frac{A}{B}$ where
$$
A=(1+u)[2+3(\sigma-u)],
$$
$$
B=(1+3u)(\sigma-u)+6(1+\sigma)u,
$$
and 
$$BA'-AB'<0.$$
Therefore, 
$$\phi_{\sigma}(u)\geq \phi_{\sigma}(\sigma)=\frac{1}{3\sigma}.$$
From this we conclude that if  (\ref{abound1}) holds, that is, if $a^2\leq\frac{1}{3\sigma}<\frac{1}{2\sigma},$ then (\ref{equal2}) holds, and hence ${\cal
Q}^a_{\sigma}$ is an invariant region for all $0<\sigma\leq 1/3,$ as claimed.  It remains to verify (\ref{abound2}) and (\ref{invQ-1}) in the case
$\sigma=1/3.$

So assume $\sigma=1/3.$ Then from (\ref{equal2}), we have
$$
\phi_{1/3}(u)=\frac{(1+u)[2+3(1/3-u)]}{(1+3u)(1/3-u)+6(1+1/3)u}.
$$
But since, $\phi_{1/3}'(u)<0$ for $0\leq u\leq 1/3,$ it follows that for every $\epsilon>0$ there exists a $\delta>0$ such that 
$$
\phi_{1/3}(u)>\frac{1}{1-\delta}
$$
in the region $0\leq u\leq 1/3-\epsilon.$  Thus, if we choose $a^2\leq \frac{1}{1-\delta},$ then (\ref{equal2}) holds, 
and so $S\geq Q_{1/3}(u)$ is an invariant region for $0\leq u\leq 1/3-\epsilon.$  It follows that (\ref{abound2}) implies that orbits can only leave the
region
${\cal Q}^a_{1/3}(\epsilon)$ through the line $u=1/3-\epsilon,$ as claimed.  The proof of Lemma \ref{lemma4} is complete.  $\Box$

\subsection{The shock speed at the Big Bang}

In this section we calculate the shock speed $s_{\sigma}(S)$ along the orbit $u=u_{\sigma}(S)$ in the limit $S\rightarrow0,$ the instant of the Big Bang.  In
Theorem \ref{theorem7} we showed that $lim_{S\rightarrow0}s_{\sigma}=\infty$ if $\sigma> 1/3,$ and the following theorem asserts that
$lim_{S\rightarrow0}s_{\sigma}=0$ for $\sigma<1/3,$ but that $lim_{S\rightarrow0}s_{\sigma}=1$ for the special value $\sigma=1/3.$  In particular, this
confirms that the estimates of Theorem \ref{theorem7} are sharp.

\begin{Theorem}\label{thm8}\label{theorem8}
Let $0<\sigma\leq 1/3,$ and let $s_{\sigma}(S)\equiv s_{\sigma}(u_{\sigma}(S))$ denote the shock speed along the solution $u=u_{\sigma}(S),$ given in
(\ref{shock}).  Then if $\sigma>1/3,$ 
\be
\label{shockspeed1a}
\lim_{S\rightarrow0}s_{\sigma}(S)=\infty,
\ee
if $\sigma<1/3,$ 
\be
\label{shockspeed1}
\lim_{S\rightarrow0}s_{\sigma}(S)=0,
\ee
and if $\sigma=1/3,$
\be
\label{shockspeed2}
\lim_{S\rightarrow0}s_{\sigma}(S)=1.
\ee  

\end{Theorem} 
\vspace{.2cm}

\noindent{\bf Proof:}  By (\ref{shock}), the shock speed is given by

\be
s_{\sigma}(S)=\frac{1}{\sqrt{S}}\left(\frac{\sigma-u_{\sigma}(S)}{1+u_{\sigma}(S)}\right),
\label{shockS}
\ee
from which (\ref{shockspeed1a}) is evident.
To verify (\ref{shockspeed1}), we show that when $0<\sigma<1/3,$ and $a=\frac{1}{\sqrt{3\sigma}},$ there exists $m>0$ such that
$$(1,0)\in R^m_{\sigma}\equiv\left\{(S,u): S\geq m(\sigma-u)\right\}\cap {\cal Q}^{a}_{\sigma},$$ and that the region $R^m_{\sigma}$ is negatively
invariant for solutions of system (\ref{1}), c.f. (\ref{invQ}).  If $R^m_{\sigma}$ is negatively invariant and contains the initial point $(S,u)=(1,0),$ then
$u_{\sigma}(S)\geq
\sigma-\frac{1}{m}S$ in a neighborhood of
$u=\sigma,$ which implies (\ref{shockspeed1}) because then we can use (\ref{shockS}) to conclude 
$$0\leq \lim_{S\rightarrow0}s_{\sigma}(S)\leq\lim_{S\rightarrow0}\frac{1}{\sqrt{S}}\left(\frac{\frac{1}{m}S}{1+u_{\sigma}(S)}\right)=0.$$

To see that $R^m_{\sigma}$ is negatively invariant for some $m>0,$ assume first that
\be
m\leq h'(\sigma)=\frac{1-3\sigma}{6\sigma(1+\sigma)},
\ee
c.f. (\ref{Gisocline}).  This implies that 
$m(\sigma-u)\leq h(u)$ in the region $0\leq u<\sigma,$ $0<S\leq1.$ Thus when $m\leq h'(\sigma),$ the region $R^m_{\sigma}$
consists of the set of all points in ${\cal Q}^a_{\sigma}$ that lie above the curve $S=m(\sigma-u)$ and below
the curve $S=h(u).$  It is also easily
verified that the initial condition $S=1,$ $u=0$ lies in $R^{m}_{\sigma}$ when $m\leq h'(\sigma).$

To find a value of $m\leq h'(\sigma)$ for which $R^{m}_{\sigma}$ is negatively invariant,  we mimic the
proof of Lemma \ref{lemma4} using
$L\equiv L(u)=m(\sigma-u)$ in place of $Q_{\sigma}(u).$  That is, $R^m_{\sigma}$ is negatively invariant if the vector field $(F,G)$ points into the
region below
$S=L(u),$ all along $S=L(u),$  which is to say that $|L'(u)|\geq
\left|\frac{dS}{du}\right|$ on
$S=L(u),$ where
$\frac{dS}{du}$ is given in (\ref{-6-1}).  Since we have already verified that ${\cal Q}^{a}_{\sigma}$ is negatively invariant for
$a=\frac{1}{\sqrt{3\sigma}},$ we actually need only verify the negative invariance of the curve $S=L(u)$ for values $\bar{u}\leq u\leq \sigma,$ where
$\bar{u}$ is the value of $u$ at the point of intersection $(\bar{S},\bar{u})$ of the line
$S=L(u)$ and the quadratic
$S=Q_a(u)=\frac{1}{3\sigma}\frac{(\sigma-u)^2}{(1+u)^2},$ c.f.  (\ref{Q}).  Since $L$ is linear and $Q_a$ is quadratic at $(0,\sigma),$ it follows that
$\bar{u}\rightarrow \sigma$ as $m\rightarrow 0.$  Thus our condition is

\be
|L'|=m\geq \frac{2L(1+3u)[L(1+u)+(\sigma-u)]}{(1+u)|(\sigma-u)(3u-1)+6u(1+u)L|},\nonumber
\ee 
for $\bar{u}\leq u\leq \sigma.$  After some algebra, this reduces to
\be
1\geq\frac{2(\sigma-u)(1+3u)[m(1+u)+1]}{|(1+u)|(3u-1)+6u(1+u)m|}.\label{invariantQ-1}
\ee
But the term inside the absolute value is negative because

\be
m\leq\frac{1-3u}{6u(1+u)},\label{detsign-1}
\ee
a consequence of the fact that we are assuming $m\leq h'(\sigma).$ 
Therefore, we can solve (\ref{invariantQ-1}) for $m$ to obtain the following condition equivalent to the condition that $S\geq L(u)$
is negatively invariant:
\be
m\leq\frac{1-3u-2(1+3u)(\sigma-u)}{(1+u)[2(1+3u)(\sigma-u)+6u]}=\psi_{\sigma}(u).\label{equal1-1}
\ee
Now assuming $\sigma<\frac{1}{3}$ is fixed, it follows that $\psi=A/B,$ where
\be\label{Bimp}
B=(1+u)[2(1+3u)(\sigma-u)+6u]\leq 2(1+\sigma)(7\sigma+3\sigma^2),
\ee
and
$$
A=1-3u-2(1+3u)(\sigma-u).
$$
Therefore, to verify (\ref{equal1-1}) for $\bar{u}\leq u\leq \sigma$ for some $m>0,$ it suffices only to show that for $m$ sufficiently small, 
$A=1-3u-2(1+3u)(\sigma-u)$ is bounded uniformly away from zero for all $u\in[\bar{u},\sigma].$  But $1-3\sigma=\epsilon>0,$ so
$\lim_{m\rightarrow0}\bar{u}=\sigma$ implies that there exists an $\delta>0$ such that $m<\delta$ guarantees that $2(1+3u)(\sigma-u)<\epsilon/2$ for all
$u\in [\bar{u},\sigma].$  In this case, $A=1-3u-2(1+3u)(\sigma-u)\geq\frac{\epsilon}{2},$ and hence by (\ref{Bimp}), 
\be
\psi_{\sigma}(u)\geq\frac{\epsilon}{4(1+\sigma)(7\sigma+3\sigma^2)}\equiv\delta_1.
\ee
It follows from (\ref{equal1-1}) that if we choose
$m=Min\{\delta,\delta_1\},$ 
Then for fixed $\sigma<\frac{1}{3},$ it follows that 
\be
m\leq\psi_{\sigma}(u),\label{equal1-2}
\ee
for all $u\in [\bar{u},\sigma],$ and hence that $R^m_{\sigma}$ is negatively invariant.  Thus the proof of (\ref{shockspeed1}) is complete, and it remains
only to consider the case $\sigma=1/3.$ 

In the case $\sigma=1/3,$  $R^m_{\sigma}$ is not negatively invariant for any $m>0,$ and to evaluate $\lim_{S\rightarrow0}s_{\sigma}(S)$ in this case
we need the asymptotics of the solution
$u_{1/3}(S)$ near
$S=0.$
This is given in the following lemma:

\begin{Lemma}\label{lemma5}

Assume $\sigma=1/3.$ Then
\be
\label{asympt3}
u_{\sigma}(S)\sim 1/3-m_*\sqrt{S},\ \ \ as\ S\rightarrow0,
\ee
where
\be
\label{asympt4}
m_*=\frac{4}{3}.
\ee
\end{Lemma}
Lemma \ref{lemma5} implies (\ref{shockspeed2}) of Theorem \ref{theorem8}, (the case $\sigma=1/3$), because using (\ref{asympt3}) in (\ref{shock}) we
obtain,

\be
s_{\sigma}(S)&=&\frac{1}{\sqrt{S}}\left(\frac{\sigma-u_{\sigma}(S)}{1+u_{\sigma}(S)}\right)\label{shockS1}\\
&=&\frac{1}{\sqrt{S}}\left(\frac{m_*\sqrt{S}}{1+u_{\sigma}(S)}\right)\rightarrow \frac{\frac{4}{3}}{1+\frac{1}{3}}=1.
\nonumber
\ee
Note that (\ref{asympt3}) confirms that the orbit $u_{1/3}(S)$ comes into the rest point $S=0,$ $u=1/3,$ asymptotically like the negatively invariant curve
$S=Q_{1}(u)=\left(\frac{3}{4}\right)^2(\frac{1}{3}-u)^2,$ c.f. (\ref{Q}).

\noindent{\bf Proof of Lemma \ref{lemma5}:}   Assume $\sigma=1/3.$  To verify (\ref{asympt3}), write
\be
u_{\sigma}(S)=1/3+\phi(S), \ \ \ 0<S\leq1.\label{phiofS}
\ee
We find the equation that $\phi$ satisfies asymptotically as $S\rightarrow0.$  (Note that $\phi(0)=0$ because $u_{\sigma}(S)$ tends to the rest point
$(0,1/3)$ as $S\rightarrow0.$)  Putting (\ref{phiofS}) into (\ref{12}) we obtain

\be
S\phi'(S)=\frac{(\frac{4}{3}+\phi)\left[-3\phi^2+(2+6\phi)(\frac{4}{3}+\phi)S\right]}{2(2+3\phi)\left[-\phi+(\frac{4}{3}+\phi)S\right]}.\label{Sphi1}
\ee
But by Lemma \ref{lemma4}, we know that ${\cal Q}^a_{1/3}$ is negatively invariant for $a=1,$ and from this it follows that
$u_{1/3}(S)$ is squeezed below the $G$-isocline and above the curve $u=Q^{-1}(S)=1/3-\sqrt{S}$ as $S\rightarrow0.$  It follows that $(1/3-u_{1/3}(S))$ is 
order $S^{1/2}$ as $S\rightarrow0.$  Using this in (\ref{Sphi1}) gives

\be
S\phi'(S)\sim\frac{(\frac{4}{3})\left[-3\phi^2+\frac{8}{3}S\right]}{2(2)\left[-\phi\right]},\label{fbest}
\ee
where $\sim$ means to leading order in $S$ as $S\rightarrow0.$  
Thus, to leading order in
$S,$ equation (\ref{fbest}) takes the asymptotic form
\be
S\phi'=\phi-\frac{8}{9}\frac{S}{\phi},\label{fbesta}
\ee
which we can write in the form
\be
(\phi^2)'=\frac{\alpha \phi^2}{S}-\beta,\label{fbest-3}
\ee
which is linear in $v=\phi^2$ with
\be
\alpha=2,\label{fbest-4}
\ee
and
\be
\beta=16/9.\label{fbest-5}
\ee
But (\ref{fbest-3}) has the general solution
\be
v=k S^{\alpha}-\frac{\beta}{1-\alpha}S,\label{solution}
\ee
where $k$ is a constant.  Since $\alpha>1,$ the first term in (\ref{solution}) is higher order in $S,$ and thus we have shown that to leading order in $S,$
equation (\ref{fbesta}) has the unique solution
\be
\phi(S)=m_*\sqrt{S},\label{solution-1}
\ee
where 

\be
m_*=\sqrt{\frac{\beta}{\alpha-1}}=4/3.
\ee
This agrees with (\ref{asympt3}) and (\ref{asympt4}).  Thus the proof of Lemma \ref{lemma4} will be complete once we prove the following lemma.
\begin{Lemma}\label{lemma6}
Assume $\sigma=1/3.$  Then every solution $u(S)$ of (\ref{12}) that enters the rest point $S=0,$ $u=1/3$ from inside the invariant region ${\cal
Q}^{a}_{1/3},$
$a=1,$ must satisfy
\be
u(S)\sim 1/3-\frac{4}{3}\sqrt{S},\ \ \ as\ S\rightarrow0.
\ee
\end{Lemma}
\vspace{.2cm}

\noindent{\bf Proof:}  By the derivation of (\ref{fbesta}), we know that 
$$
u(S)=1/3+\hat{\phi}(S),
$$
where $\hat{\phi}$ satisfies
\be
\left(\hat{\phi}^2\right)'=\alpha\frac{\hat{\phi}^2}{S}-\beta-\epsilon(S),\label{epsilon-1}
\ee
for some function, $\epsilon(S)\rightarrow0$ as $S\rightarrow0.$ Thus it suffices to show that
$\lim_{S\rightarrow0}\left|\frac{\hat{\phi}^2(S)-\phi^2(S)}{S}\right|=0.$ But (\ref{epsilon-1}) is linear and has the general solution
\be
\hat{\phi}^2(S)=KS^{\alpha}+S^{\alpha}\int_{S}^{S_0}\left\{\beta+\epsilon(t)\right\}t^{-\alpha}dt,\label{epsilon-2}
\ee
for some constants $K$ and $S_0.$  Moreover, $\phi^2(S)=m_*^2 S$ also satisfies
\be
\phi^2(S)=\frac{\beta }{\alpha-1}S_0^{-\alpha+1}S^{\alpha}+S^{\alpha}\int_{S}^{S_0}\beta t^{-\alpha}dt,
\ee    
as one easily sees by integration, and consistent with the fact that $\phi$ solves (\ref{epsilon-1}) with $\epsilon=0.$  Thus

\be
\hat{\phi}^2-\phi^2&=&-\frac{\beta}{\alpha-1}S_0^{-\alpha+1}S^{\alpha}+S^{\alpha}\int_{S}^{S_0}\epsilon(t)t^{-\alpha}dt+KS^{\alpha}\nonumber\\
&=&-\frac{\beta}{\alpha-1}S_0^{-\alpha+1}S^{\alpha}+\epsilon_*S^{\alpha}\int_{S}^{S_0}t^{-\alpha}dt+KS^{\alpha}\nonumber\\
&=&-\frac{\beta}{\alpha-1}S_0^{-\alpha+1}S^{\alpha}+\epsilon_*\left(\frac{1}{1-\alpha}\right)\left[S_0^{-\alpha+1}S^{\alpha}-S\right]+KS^{\alpha},\nonumber\\\label{difff0}
\ee
where we have applied the mean value theorem for integrals with $\epsilon_*=\epsilon(S_*)$ for some $S_*\in(S,S_0).$  Now if we choose $S_0=S^{a}$ for some
$0<a<1,$ then every term in (\ref{difff0}) is higher order than $S.$  That is,
\be
\left|\hat{\phi}^2-\phi^2\right|(S)\leq\left[\frac{\beta+\epsilon_*}{\alpha-1}S^{(\alpha-1)(1-a)}+
\left|K\right|S^{\alpha}+\frac{\epsilon_*}{\alpha-1}\right]S,\nonumber
\ee
from which we conclude that
\be
\lim_{S\rightarrow0}\left|\frac{\hat{\phi}^2(S)-\phi^2(S)}{S}\right|=0,\nonumber
\ee
as claimed.  This completes the proof of Lemma \ref{lemma6}, and thus Lemma \ref{lemma5} as well. 
$\Box$

\subsection{The asymptotics for $S\geq1$}

Equation (\ref{Sarrow1}) implies that $v\equiv\frac{\bar{\rho}}{\rho}=0=\bar{\rho}$
at
$S=1.$ Thus, the TOV metric {\it inside the Black Hole} continues to the empty space Schwarzschild metric at $u=0,$ $v=0,$ $S=1,$ an event horizon for the
outer TOV metric in light of the fact that at $S=1,$ $N=\frac{2M}{\bar{r}}=1,$ c.f. (\ref{shockdistance}).
 It follows that if the FRW density
$\rho$ is small at
$S=1,$ then it makes sense in cosmology to approximate the FRW solution with
$p=0$ for all times
$S\geq1.$ Assuming this, Theorem
\ref{theorem6} implies that for each
$0<\sigma<1,$ the unique solution
$u_{\sigma}(S)$ continues to the zero pressure,
$k=0,$ OS solution at
$S=N=1,$ the moment when the shock wave lies exactly one Hubble length from the FRW center.  That is, at $S=1,$ the shock wave emerges from the White Hole
event horizon of an ambient Schwarzschild metric as an outward propagating contact discontinuity that bounds a finite FRW mass, this being exactly equal
to the total mass of the ambiant Schwarzschild metric into which it propagates.
Thereafter the interface continues out to infinity along a geodesic of the Schwarzschild metric outside the Black Hole. 

We conclude that the OS solution gives the
large time asymptotics of this new class of shock wave solutions that evolve inside of a Black Hole---and thus the explosion that begins at the Big
Bang eventually settles down to a localized expansion that looks something like a giant supernovae, but on an enormous scale. .  
\vspace{.2cm}

\section{Estimates for the Shock Position}\label{sect7}
\setcounter{equation}{0}
In this section we use the invariant region ${\cal Q}^a_{\sigma},$ (c.f. (\ref{invQ})), to estimate the shock position
$\bar{r}_0$ at present time in terms of its position at the instant of the Big Bang $S\equiv 1/N=0.$  Using this, we finish by estimating the time at which
the shock wave first leaves the Black Hole in terms of the time at which the shock wave first becomes visible to an observer at the FRW center.   

Since the physical shock position at
$S=0$ is
$\bar{r}=0,$ we begin by estimating
$\bar{r}_0=r_0,$ $R(t_0)=1,$ in terms of $r_*,$ the value of the radial FRW coordinate at the instant of the Big Bang.  In particular, the analysis shows
that 
$r_{\sigma}(S)$ has a limit $r_*\equiv r_*(\sigma)$ as $S\rightarrow0,$ for any solution
$(r_{\sigma}(S),u_{\sigma}(S))$ of system (\ref{ueqn}),(\ref{Neqn}).  
So assume the
FRW solution for the equation of state $p=\sigma\rho,$ $0<\sigma\leq 1/3,$ is given, together with the shock trajectory $u_{\sigma}(S),$ that solves equation
(\ref{ueqn}).  There then remains the one equation (\ref{Neqn}), leaving one free initial condition to impose. 
The next lemma gives an equivalent form of equation (\ref{Neqn}) in terms of the FRW variable $r.$

\begin{Lemma}
\label{lemmareqn}
Equation (\ref{Neqn}) is equivalent to 

\be
\label{eqnr}
\frac{dr}{dS}=\frac{\sigma-u}{(1+\sigma)(1+3u)}\frac{r}{S},
\ee
for all $0<S\leq 1.$
\end{Lemma}

\vspace{.2cm}
\noindent{\bf Proof:}  Starting with (\ref{Neqn}), we can write

\be
\frac{d\bar{r}}{dS}=\frac{d\bar{r}}{dN}\frac{dN}{dS}=\frac{1}{(1+3u)}\frac{\bar{r}}{S}.
\label{a.1}
\ee
Also, 

\be
\frac{dr}{dS}=\frac{d\left(\frac{\bar{r}}{R}\right)}{dS}=\frac{1}{R}\frac{d\bar{r}}{dS}-\frac{\bar{r}}{R^2}\frac{dR}{dS},
\label{a.1.1}
\ee
so using (\ref{a.1}) we have

\be
\frac{dr}{dS}=\left(\frac{1}{1+3u}\right)\frac{r}{S}-\frac{r}{R}\frac{dR}{dS}.
\nonumber
\ee
So now we need only compute $dR/dS.$  For this, we use (\ref{frw4-1}), (\ref{frwH}) to obtain

\be
H=H_0R^{-\frac{3(1+\sigma)}{2}},
\nonumber
\ee
and from (\ref{interpretN}),

$$
H^2\bar{r}^2=N=1/S.
$$
These imply

\be
R=\left(SH_0^2\bar{r}^2\right)^{\frac{1}{3(1+\sigma)}}.
\label{a.2}
\ee
Using (\ref{a.2}) together with (\ref{a.1}) gives

\be
\frac{dR}{dS}=\frac{(1+u)}{(1+\sigma)(1+3u)}\left(H_0^2\bar{r}^2\right)^{\frac{1}{3(1+\sigma)}}S^{-\frac{(2+3\sigma)}{3(1+\sigma)}}.
\label{a.3}
\ee
From (\ref{a.2}) we get

\be
H_0^2\bar{r}^2=\frac{R^{3(1+\sigma)}}{S},
\nonumber
\ee
and using this in (\ref{a.3}) gives

\be
\frac{dR}{dS}=\frac{(1+u)}{(1+\sigma)(1+3u)}\frac{R}{S}.
\label{a.4}
\ee
Use this in (\ref{a.1.1}) to get (\ref{eqnr}). $\Box$

We now use (\ref{eqnr}) to estimate $r_0$ in terms of $r_*,$   assuming that
the shock position $\bar{r}_0=r_0=\sqrt{N_0}/H_0$ at present time lies beyond one Hubble length.  Then $N_0>1,$ and we have

\be
S_0=1/N_0=\frac{1}{H_0\bar{r}_0^2}<1,
\label{S0lessthanone}
\ee 
c.f. (\ref{frw1}), (\ref{shockdistance}).  Then integrating equation (\ref{eqnr}) we obtain,

\be
\label{formulaforrbar}
r_0=r_*e^{\int_{0}^{S_0}\left(\frac{\sigma-u}{1+3u}\right)\frac{1}{(1+\sigma)S}dS},
\ee
where $u$ denotes the function of $S$ given by the trajectory $u= u_{\sigma}(S))$ of (\ref{ueqn}), $u_{\sigma}(0)=\sigma\leq 1/3,$ $u_{\sigma}(1)=0,$
$u'_{\sigma}(S)<0,$ $0\leq S\leq 1.$
We can now use the invariant region ${\cal Q}^a_{\sigma}$ to estimate $\sigma-u_{\sigma}(S)$ for $0<S\leq1.$  That is, the condition that the orbit
$u_{\sigma}(S)$ lies in 
${\cal Q}^a_{\sigma}$ implies that

\be
Q_a(u_{\sigma}(S))\equiv a^2\frac{(\sigma-u)^2}{(1+u)^2}\leq S\leq Min\left\{1,h(u_{\sigma}(S)\right\},
\label{ineqalS}
\ee
holds all along the orbit,
where
\be
\label{hest}
h(u)=\frac{(\sigma-u)(1-3u)}{6u(1+u)}\leq\frac{(\sigma-u)(1-3u)}{6u}\leq\frac{\sigma-u}{6u},
\ee
and we can take, c.f. (\ref{abound1}),
\be
\label{aest}
a^2=\frac{1}{3\sigma}.
\ee
Now using (\ref{aest}) in the lower bound in (\ref{ineqalS}) leads to

\be
\label{lowerbound}
\frac{\sigma-u}{1+3u}\leq \left(\frac{1+u}{1+3u}\right)\sqrt{3\sigma}\sqrt{S}\leq \sqrt{3\sigma}\sqrt{S},
\ee
and applying this in (\ref{formulaforrbar}) gives

\be
r_0\leq r_*e^{\frac{2\sqrt{3\sigma}}{1+\sigma}\sqrt{S_0}}.
\label{lowerbound1}
\ee

We now apply the upper bound in (\ref{ineqalS}). Note first that

\be
\label{upperbound1}
Min\left\{1,h(u_{\sigma}(S)\right\}\leq Min\left\{1,\frac{(\sigma-u)(1-3u)}{6\hat{u}}\right\}\leq Min\left\{1,\frac{\sigma-u}{6\hat{u}}\right\},
\ee
where $\hat{u}$ is the (smallest) value of $u$ at which $h(\hat{u})=1.$  (We need $\hat{u},$ the smallest value of $u$ that puts $h(u)\leq1,$ to bound the
factor
$u$ in the denominator of
$h(u).$) We can estimate $\hat{u}$ as follows.  First, note that by (\ref{hest}),
$h(\hat{u})=1$ is equivalent to

$$
\hat{u}=\frac{\sigma}{7+3\sigma}-\frac{3\hat{u}^2}{7+3\sigma}\leq \frac{\sigma}{7+3\sigma},
$$ 
so using the latter inequality to estimate the second term after the equality, we obtain

\be\label{hatu}
\hat{u}&\geq&\frac{\sigma}{7+3\sigma}-\frac{3}{7+3\sigma}\left(\frac{\sigma}{7+3\sigma}\right)^2\nonumber\\
&\geq&\frac{1-1/7^2}{8}\sigma\geq \frac{\sigma}{9}.
\ee
Using (\ref{hatu}) in the second inequality in (\ref{upperbound1}), (i.e., ignoring for the moment the factor $(1-3u)$ in the middle term of
(\ref{upperbound1})),  implies that

\be
\label{upperbound2}
S\leq Min\left\{1,\frac{\sigma-u}{6\hat{u}}\right\}\leq Min\left\{1,\frac{3}{2}\frac{\sigma-u}{\sigma}\right\},
\ee
which implies that

\be
\label{upperbound3}
\sigma-u\geq \frac{2}{3}\sigma S
\ee
 holds all along the orbit $u=u_{\sigma}(S).$  Using (\ref{upperbound3}) in (\ref{formulaforrbar}) gives the inequality

\be
\label{upperbound4}
\bar{r}_0\geq r_*e^{\frac{1}{4}\sigma S_0}.
\ee 
In the case $\sigma=1/3,$ we can improve the estimate (\ref{upperbound4}) by using the first inequality in (\ref{upperbound1}) to obtain
\be
\label{upperbound2-1}
S\leq Min\left\{1,\frac{(\sigma-u)^2}{18\hat{u}}\right\}\leq Min\left\{1,\frac{1}{2}\frac{(\sigma-u)^2}{\sigma}\right\},
\ee
and

\be
\label{upperbound3-1}
\sigma-u\geq \sqrt{2\sigma S},
\ee
in place of (\ref{upperbound2}) and (\ref{upperbound3}). Using (\ref{upperbound3-1}) in (\ref{formulaforrbar}) gives the improved inequality valid for
$\sigma=1/3,$
\be
\label{upperbound4-1}
\bar{r}_0\geq r_*e^{\frac{\sqrt{6}}{4}\sqrt{S_0}}.
\ee

Putting (\ref{lowerbound1}) and (\ref{upperbound4}) together, we obtain the following bounds for the shock position
$r_0=\bar{r}_0$ in terms of the initial position $r_*$ that apply for $0<\sigma\leq 1/3:$

\be
\label{mainboundsonposition}
r_*e^{\frac{1}{4}\sigma S_0}\leq r_0\leq r_*e^{\frac{2\sqrt{3\sigma}}{1+\sigma} \sqrt{S_0}}.
\ee
In the case $p=\frac{1}{3}\rho,$ we obtain the improved bounds

\be
\label{mainboundsonposition-1}
r_*e^{\frac{\sqrt{6}}{4}\sqrt{S_0}}\leq r_0\leq r_*e^{\frac{3}{2}\sqrt{S_0}}.
\ee
The following Corollary follows immediately from (\ref{mainboundsonposition}) and (\ref{mainboundsonposition-1}) in the case $S_0=1:$ 

\begin{Corollary}
Let $\bar{r}=\bar{r}_{crit}$ denote the FRW shock position at the instant $S=1$ when the shock wave emerges from the Black Hole. Then for $0\leq \sigma\leq
1/3$ we have
\be
r_*e^{\frac{1}{4}\sigma}\leq \bar{r}_{crit}\leq r_*e^{\frac{2\sqrt{3\sigma}}{1+\sigma}},
\label{eqnr1}
\ee
while if $\sigma=1/3$ we have,
\be
\label{mainboundsonposition-2}
r_*e^{\frac{\sqrt{6}}{4}}\leq \bar{r}_{crit}\leq r_*e^{\frac{3}{2}},
\ee
where $r_*$ is the FRW radial coordinate of the shock wave at the instant of the Big Bang.
\end{Corollary}

\noindent Note that because

\be
\label{estimateS0}
S=\frac{1}{N}=\frac{1}{H^2\bar{r}^2},
\ee
multiplying (\ref{mainboundsonposition}) through by $H_0$ gives

\be
\label{mainboundsonposition1}
\frac{e^{-\frac{2\sqrt{3\sigma}}{1+\sigma} \sqrt{S_0}}}{H_0r_*}\leq \sqrt{S_0}\leq \frac{e^{-\frac{1}{4}\sigma S_0}}{H_0r_*}\leq \frac{1}{H_0r_*},
\ee
and so we could use (\ref{mainboundsonposition1}) in (\ref{mainboundsonposition}), (\ref{mainboundsonposition-1}) to obtain estimates involving $H_0$ in place
of
$S_0.$  In particular we have

\be
\label{mainboundsonposition2}
r_*\leq r_0\leq r_*e^{\frac{2\sqrt{3\sigma}}{(1+\sigma)H_0r_*}},
\ee
which reproduces the OS result $r_0=r_*$ in the limit $\sigma\rightarrow0.$  

The final theorem gives an estimate
for the number of Hubble lengths to the shock wave at the instant when it first becomes visible at the FRW center, as well as an estimate for the time it
takes the shock wave to emerge from the Black Hole after it first becomes visible at the FRW center.

\begin{Theorem}\label{theorem9}
Let $r_*=\lim_{S\rightarrow0}r_{\sigma}(S)$ denote the FRW position of the shock wave at the instant of the Big Bang, and assume $0<\sigma\leq 1/3.$   Then
the shock wave will first become visible at the center $\bar{r}=0$ of the FRW spacetime at FRW time $t=t_0,$ at the moment when the Hubble constant
$H_0=H(t_0)$ satisfies

\be
H_0r_*=\frac{2}{1+3\sigma},\label{frstest}
\ee
(assuming $R=1$ at $t=t_0$),  and, the number of Hubble lengths $\sqrt{N_0}$ from the FRW center
to the shock wave at time $t=t_0$ satisfies

\be
\label{finalestimate1}
1\leq\frac{2}{1+3\sigma}\leq \sqrt{N_0}\leq \frac{2}{1+3\sigma}e^{\sqrt{3\sigma}\left(\frac{1+3\sigma}{1+\sigma}\right)}.
\ee 
Furthermore, the time $t_{crit}>t_0$ at which the shock wave will emerge from the Black Hole given that it first becomes visible at $t=t_0$ is estimated by

\be
\label{finalestimate2-11}
\frac{2}{1+3\sigma}e^{\frac{1}{4}\sigma}\leq \frac{t_{crit}}{t_0}\leq
\frac{2}{1+3\sigma}e^{\frac{2\sqrt{3\sigma}}{1+\sigma}},
\ee
and by the better estimate
\be
\label{finalestimate2-22}
e^{\frac{\sqrt{6}}{4}}\leq \frac{t_{crit}}{t_0}\leq e^{\frac{3}{2}},
\ee
in the case $\sigma=1/3.$
\end{Theorem}

\noindent Note that, for example, (\ref{finalestimate1}), (\ref{finalestimate2-11}) imply that at the OS limit $\sigma=0,$ 
$$\sqrt{N_0}=2,\ \  \frac{t_{crit}}{t_0}=2,$$ 
and in the limit $\sigma=1/3,$
$$1.8\leq\frac{t_{crit}}{t_0}\leq 4.5,\ \ 1<\sqrt{N_0}\leq 4.5.$$
Note also that (\ref{finalestimate1}) and (\ref{finalestimate2-11}) imply that the shock wave will still lie beyond one Hubble length at the time $t=t_0$
when it first becomes visible at the FRW center. 
\vspace{.2cm}

\noindent{\bf Proof of Theorem \ref{theorem9}:} Equation (\ref{inftyexact}) implies that
if the shock wave is first visible at
$t=t_0,$
$R_0=1,$ then 

\be
\label{rstarinf}
r_*=r_{\infty}=\frac{2}{(1+3\sigma)H_0},
\ee
which implies (\ref{frstest}).  To verify  (\ref{finalestimate1}), multiply equation
(\ref{mainboundsonposition2}) through by
$H_0,$ and use (\ref{frstest}) and (\ref{shockdistance}).

To verify (\ref{finalestimate2-11}) and (\ref{finalestimate2-22}), let $\bar{r}_{crit}=r_{crit}R_{crit}$ denote the shock position at $S=N=1,$ (the instant
when the shock wave emerges from the Black Hole), and use (\ref{frwH}) and (\ref{estimateS0}) to write

\be
\label{rstarinf1}
\frac{t_{crit}}{t_0}=\frac{H_0}{H_{crit}}=H_0\bar{r}_{crit}.
\ee
Then multiplying (\ref{eqnr1}) and (\ref{mainboundsonposition-2}) through by $H_0$ and using (\ref{rstarinf}) in (\ref{rstarinf1}) gives
(\ref{finalestimate2-11}) and (\ref{finalestimate2-22}), respectively.  This completes the proof of Theorem \ref{theorem9}.
$\Box$


\section{Concluding Remarks}
\setcounter{equation}{0}

We have constructed global exact solutions of the Einstein equations in which the expanding FRW
universe extends out to a shock wave that lies arbitrarily far beyond the Hubble length.  The critical OS solution {\em inside the Black Hole} is obtained
in the limit of zero pressure, but the shock wave solutions have qualitative differences from the OS solution.  For example,
the shock surface
$\bar{r}(t)$ tends to zero as $t\rightarrow0,$ and the mass function at the shock $M(\bar{r}(t))$ is finite for all
$t>0,$ but unlike the OS solution, in the shock wave solution,
$M(\bar{r}(t))$ tends to infinity as $t\rightarrow0.$ That is, the mass function is infinite at the
instant of the Big Bang, but immediately becomes a finite decreasing function of FRW time, for all future times
$t>0.$   Moreover, when $p\neq0,$ the directional orientation of the shock wave
motion relative to the various observers is determined by the entropy condition--the entropy condition chooses the explosion over the implosion.   For the
entropy condition we take the condition that the pressure and density be larger behind the shock wave; that is, larger on the side that receives the mass
flux.  This condition implies that the shock is compressive, and is sufficient to rule out expansion shocks in
classical gas dynamics, \cite{smol}.

One can ask the question, what is the solution like beyond the shock wave at any fixed instant of time inside the Black Hole?  The answer is
that the TOV energy density $\bar{\rho}(\bar{r})$ and the TOV mass $M(\bar{r})$ as well, are both
constant at each fixed \lq\lq time\rq\rq in the TOV metric beyond the shock wave, because $\bar{r}$ is the timelike TOV coordinate {\em inside the Black
Hole}.  This is no contradiction because what we identify (via shock matching) as the total mass function on the TOV side of the shock comes
from the
$d\bar{r}^2$ component of the metric, which is {\em timelike} inside the Black Hole.  However, this TOV \lq\lq total mass\rq\rq  matches the FRW total
mass continuously at the shock surface, and the FRW total mass has the physical interpretation as $M=\frac{4\pi}{3}\rho\bar{r}^3,$ the integral of the
energy density at each fixed time $t$ in the FRW coordinates $(t,r).$  \footnotemark[15]\footnotetext[15]{To see how $M$ can be constant on the TOV side when
it measures a total mass on the FRW side, consider shock matching in
$(t,\bar{r})$ coordinates.  Then since the FRW mass $M$ depends only on $t,$ while the TOV mass depends only on $\bar{r},$ one of them depends on the
timelike coordinate and one on the spacelike coordinate in the $(t,\bar{r})$ coordinates at the shock. Thus, in the Einstein equations, the nonzero $M$
derivative ends up in a different equation on each side of the shock, giving the physical total mass on the FRW side.   But on the TOV side, the derivative
$M'$ is equated to the pressure, and thus doesn't have the same interpretation as an integral of the energy density.}  Thus the evolution of the ``total
mass\lq\lq is interesting and surprising {\em inside the Black Hole}.  

Throughout its expansion, the strength of the shock is on the
order of the energy density $\rho$ on the FRW side of the shock.  At the moment when the shock wave lies at the critical distance of exactly one Hubble length
from the FRW center, the TOV density and pressure are zero, and thus we argue that if the FRW density is small as well, then the shock wave continues, (with
small errors), to a zero pressure OS interface leaving the Black Hole at that instant. Thus the OS solution provides the large time asymptotics
of these shock wave models.  That is, the interface that marks the boundary of the FRW expansion continues out through the White Hole event horizon of an
ambient Schwarzschild metric at the instant when the shock wave is exactly one Hubble length from the FRW center
$\bar{r}=0,$ and it then continues on out to infinity along a geodesic of this Schwarzschild metric outside the Black Hole. 
These solutions thus indicate a scenario for the Big Bang in which the expanding universe emerges from an explosion
emanating from the White Hole singularity inside the event horizon of an asumptotically flat Schwarzschild spacetime of
finite mass.  The model does not require the physically implausible assumption that the uniformly expanding portion of the
universe is of infinite mass and extent at every fixed time, and it has the nice feature that it embeds the Big Bang
singularity of cosmology within a larger spacetime, the Schwarzschild spacetime.  Moreover, the model
also allows for arbitrarily large densities to exist over arbitrary numbers of Hubble lengths early on in the Big Bang, a
prerequisite for the standard physics of the Big Bang at early times.  

One might ask how an observer near the FRW center would first detect evidence of such a cosmic shock wave.  Since the shock wave emerges from the
Big Bang beyond the Hubble length, the model would imply a uniform expansion throughout a region that is initially well beyond the light cone of an observer
positioned near the FRW center.  If the shock wave were initially far enough out, then the uncoupling of matter from radiation at about $300,000$ years
after the Big Bang would produce an extended region with a uniform background radiation field.  This would persist until roughly the time when the
Hubble length catches up to the shock wave, a time determined by the initial conditions.   The influence of the solution beyond the shock wave
would propagate into this radiation field at the speed of light, first appearing to an observer that is off center on the FRW side of the shock as a
disturbance in the background radiation field at a point in the sky in the direction nearest the shock wave, and the disturbance would grow from
that time onward.

These exact shock wave solutions give the global dynamics of strong
gravitational fields in an exact solution, the dynamics is qualitatively different from the dynamics of solutions when the
pressure
$p\equiv0,$ and the solution suggests a Big Bang cosmological model in which the expanding universe is bounded throughout its
expansion.  Surprisingly, unlike shock matching outside the Black Hole, the equation of state $p=\frac{1}{3}\rho$ of early Big Bang physics, plays a special
role in the equations, and for this equation of state alone, the behavior of the shock wave at the instant of the Big Bang is distinguished.
But these solutions are only rough qualitative models because the equation of state on the TOV side is determined by the equations, and therefore
cannot be imposed. That is, the TOV density
$\bar{\rho}$ and pressure
$\bar{p}$ only satisfy the loose physical bounds
$0<\bar{p}<\bar{\rho}$\ ; and on the FRW side, the equation of state is taken to be
$p=\sigma\rho,$ $\sigma\equiv const.,$ $0<\sigma<1.$  We
take these bounds as implying that the equations of state are qualitatively reasonable.   The entropy condition,
$\rho>\bar{\rho},$ $p>\bar{p},$ (that the density and pressure be larger on the side that receives the mass flux), implies
that the shock wave is compressive, and this fixes a time orientation for solutions, and determines a unique solution.\footnotemark[15]
\footnotetext[15]{ The time orientation of a
solution must be selected based on an extra condition, such as an entropy condition for shocks, because the Einstein
equations and the compressible Euler equations, taken by themselves, are both time reversible, \cite{lax,smol}.  Thus the
entropy condition for the shock is what determines the time orientation for the global dynamics of the solutions we
construct: the FRW metric expanding outward behind a shock wave emanating from a White Hole is entropy satisfying, while its time reversal, the
FRW metric contracting into a Black hole, is entropy violating.  We find it interesting that the entropy condition determines a unique solution in
the large.}   However, we expect that these solutions will capture the gross dynamics arising when more general equations of state are imposed.  In fact, we
suggest  that the global dynamics described in these solutions could {\em only} be discovered within a class of exact
solutions in which simplifying assumptions are made. For more general equations of state, other waves, (e.g. rarefaction
waves), would need to be present to meet the conservation constraint, and thereby mediate the transition across the shock
wave.   Such transitional waves would be pretty much impossible to model in an exact solution.   

Finally, we note that because Einstein's theory by itself does not choose an
orientation for time,  it
follows that if we believe that a Black Hole can exist in the forward time collapse of a mass through an event horizon as
$t\rightarrow\infty,$ (the time $t$ as observed in the far field), then we must also allow for the possibility of the time
reversal of this process, a White Hole explosion of matter through an event horizon coming from
$t\rightarrow -\infty.$   These solutions might be relevant in explaining astropysical systems, such
as galaxies and stellar associations \cite{burbhona}, whose expansions appear so great as to have emerged from an event
horizon at earlier times--an impossibility if one only allows the time orientation of a collapsing Black Hole, and not its
time reversal, the expanding White Hole.  Of course, this naturally leads one to
wonder if there is a connection between the mass that mysteriously disappears into Black Hole singularities,
and the mass that mysteriously emerges from White Hold singularities.

\end{document}